\documentclass[3p,times]{elsarticle}
\usepackage{ecrc}
\volume{00}
\firstpage{1}
\journalname{}

\runauth{}
\jid{procs}
\jnltitlelogo{}
\usepackage{amssymb}
\usepackage{amsthm}

\usepackage{amsthm}   
\usepackage{setspace}
\usepackage[figuresright]{rotating}

\usepackage{setspace}   
\usepackage{booktabs}   
\usepackage{amsmath}
\usepackage{amssymb}
\usepackage[tight]{subfigure}

\newtheorem{thm}{Theorem}

\newcommand{\mc}{\mathcal}
\newcommand{\mb}{\mathbf}

\newcommand{\mbb}{\mathbb}

\begin{document}
\begin{frontmatter}

\dochead{}

\title{A Scaling Law for Short Term Load Forecasting on Varying Levels of Aggregation}

\author{
Raffi Sevlian and Ram Rajagopal \thanks{R. Sevlian is with the Department of Electrical Engineering and the Stanford Sustainable Systems Lab, Department of Civil and Environmental Engineering, Stanford University, CA, 94305. Email: rsevlian@stanford.edu.}%
\thanks{R. Rajagopal is with the Stanford Sustainable Systems Lab, Department of Civil and Environmental Engineering, Stanford University, CA, 94305. 
R. Rajagopal is supported by the Powell Foundation Fellowship. Email: ramr@stanford.edu.}
}

\address{}

\begin{abstract}
This paper proposes a simple empirical scaling law that describes load forecasting accuracy at varying levels of aggregation.  
We show that for many forecasting methods, aggregating more customers improves the relative forecasting performance up to specific point. 
Beyond this point, no more improvement in relative performance can be obtained.  
A benchmarking procedure for applying the scaling law to different forecasting models is presented. 
The aggregation model is evaluated with year long consumption profiles of over 180 thousand Pacific Gas \& Electric customers.
A theoretical model based on a bias variance decomposition of the forecast error is used to model the \textit{Aggregation Error Curves} (AECs) that are empirically explored.
\end{abstract}

\begin{keyword}
\end{keyword}

\end{frontmatter}

\section{Introduction}
\label{section-introduction}

To meet the challenges posed by a significant increase in distributed energy resources, there is a growing need for active control at the distribution level.
The deployment of improved sensing and control technologies allow a variety of applications in the distribution system.
As an example, smart meter data has been proposed to be used in various planning and operational applications.
In applications such as topology processing or state estimation, load forecasts of 1 hour up to a day ahead are needed to provide pseudo-measurements.
For the design of the next generation of distribution system applications, understanding the variability of these pseudo-measurements is of importance.
The focus of this paper is the following: relying on empirical analysis and theoretical modeling we develop an intuitive scaling law for load forecasts on varying levels of aggregation.

The field of load forecasting is very mature with numerous methodologies having been proposed throughout the years.  
These works have focused primarily on the level of large substations servicing tens of megawatts or up to an entire transmission system which has a load of tens of gigawatts. 
With recent advances in communication infrastructure for remote measurement and automated metering, there is an abundance of new data from homes and commercial buildings.  
The proliferation of this more granular data has led to an increase in forecasting research at these lower levels aggregation.  
Typical home loads are $1$ to $2$ kWh, while commercial buildings can be $100$ times that amount.  
The relative forecasting errors typically seen at the level of substations and power systems has been quite low ($1\%-2\%$) while forecasting performance at the individual level show much higher errors (up to $30\%$). 
    
Clearly, there is a discrepancy in forecasting performance at the level of individuals and of the entire power grid.  
This is the \textit{effect of load aggregation} on forecasting performance.  
Our work aims to quantify the effect of aggregation by proposing a scaling law relating kWh load level and forecasting error.  
Using data from close to two hundred thousand residential customers and commercial buildings, we construct datasets of varying aggregation levels and verify a proposed aggregation model via an \textit{Aggregation Error Curve}.

A theoretical aggregation scaling law is obtained by assuming a simple underlying consumption model for each individual then decomposing the aggregate error into bias and variance terms.  
The theoretical scaling law is then fit to the experimental AEC for various forecasting methods, and horizons.
It is shown that the scaling law holds for a wide range for forecasting methods, prediction horizons and data types.

We propose that this AEC can be used to benchmark various proposed forecasting procedures in smart grid applications.
Since any load forecasting done for smart grid applications on the distribution system will require small to moderate aggregation of loads as the basic unit analysis, having an AEC is helpful to understand the performance of a forecasting method on a wide yet practical range of loads that will be encountered in practice.
We should also note that while using off the shelf forecasting techniques with aggregation can improve forecasting accuracy, development of custom finely tuned forecasting methods will almost always outcompete this proposed strategy.
\textit{This benchmarking is merely a tool to understand roughly how forecasters should behave in different aggregation levels, and not make claims against any specific forecasting method.}

The paper is organized as follows. 
Section \ref{section-Literature-Review} provides a review of energy forecasting at short time scales.
Section \ref{section-Modeling-Load-Aggregation} develops a model for aggregation and proposes a set of scaling laws for two common forecasting metrics.  
The scaling laws are then verified in Sections \ref{section-Experiment-Setup} and \ref{section:Experimental-Results}.

\section{Literature Review}   
\label{section-Literature-Review}
Here we survey both general methods in short term forecasting in Section \ref{subsection-Short-Term-Load-forecasting} as well as specific work which work in forecasting intermediate sized aggregates.
We refer to load forecasting as point forecasts of expected value and not probabilistic forecasts.
A review of probabilistic forecasting is done in \cite{Hong2016}.

\subsection{Short Term Load forecasting}
\label{subsection-Short-Term-Load-forecasting}
A general overview of short term load forecasting state of the art is provided in \cite{Hong2014}, \cite{HongPHD}; and more classic surveys are given in \cite{gross1987short}, \cite{weron2007modeling} and \cite{harris2011electricity}.
We review a few popular methods that we utilize as procedures in this paper. 
Seasonal ARMA and other linear modeling approaches are considered in \cite{Papalexopoulos1990}.
Seasonal Vector modeling with segment identification is considered first in \cite{Espinoza2005}.
Neural networks have been applied to load forecasting for quite some time.  
Some early papers are  \cite{lee1992short}, \cite{zhang1998forecasting}, \cite{bakirtzis1996neural}, \cite{lu1993neural}.
In \cite{Hippert2001}, the author provides a comprehensive survey of the current state of Neural networks applied to load forecasting.
Support vector regression has recently been applied to load forecasting as well with substantial work done by \cite{Pai2005A} and \cite{Pai2005B}.

\subsection{Hour Ahead Forecasting on Individual and Large Aggregates}
\label{subsection-Short-Term-Load-forecasting}
Recent work shows a fundamental limitation to the predictability of individual customers.
\cite{Edwards2012} performs one hour ahead forecasting based on hourly data utilizing machine learning. 
The methods achieve a relative error of $1.61\%$ to $13.41\%$ for a $700$ kWh commercial building and between $15\%$ to  $30\%$ for three homes with mean consumption close to $1.5$ kWh.  
In \cite{Ziekow2013}, machine learning methods are compared on data from three homes with mean consumption $1$ to $2$ kWh achieving relative errors close to  $25\%$. 
In \cite{Singh2012}, various methods are utilized to forecast peak demand for individual homes. 
The authors conclude that seasonal autoregressive models achieve the best performance, with relative error of $30\%$.   
 In \cite{Ghofrani2011}, a Kalman filter based forecaster is applied to single home data with mean consumption of $0.8$ kWh to achieve an error of $30\%$. 

Low relative errors are reported at high aggregation levels.
In \cite{Adepoju2007} the authors use an artificial neural network to forecast a mean load of $2.5$ GWh, with errors ranging from $1.73 \%$ to $3.02 \%$. 
In \cite{Benaouda2006} the authors apply wavelet multi-scale decomposition based autoregressive approaches.  
They report values of $0.7\%$ to $3.5\%$ depending on the method used on a dataset with mean load $9$ GWh. 
In \cite{Senjyu2002} artificial neural networks are applied to data with a mean consumption of $800$ MWh achieving errors from  $1.11\%$ to $1.63\%$.  
Similarly,  \cite{AlShareef2008} obtains an error in the range $0.81\%$ to $1.21\%$ utilizing neural networks on a $8$ GWh load data. 
 In \cite{Drezga1999} a novel ANN architecture is applied to two datasets with peak load $4.4$ GWh and report error between $0.8\%$ and $1.5\%$. 
 Finally, \cite{lee1992short} applies artificial neural networks to attain an error rate of $1.7\%$  for a load of $7$ GWh.
 In the experimental comparisons in this paper we benchmark hour ahead forecasting of one hour, multiple hour and day ahead intervals of consumption.
 
\subsection{Recent Work on Aggregation Forecasting}
\label{subsection-Recent-Work-on-Aggregation-Forecasting}
Initial work on developing a model for forecasting accuracy with an explicit scaling law was done by the authors in \cite{Sevlian2013}.
The model was limited to 2,000 residential customers and was unable to capture scaling behavior at large aggregate levels. 
Other work also rely on small datasets and show similar results as in \cite{Sevlian2013}.
In \cite{Mirowski2014} the authors aggregate up to 1000 customers for one hour ahead forecasting and provide a qualitative rationale for the effect.
In \cite{Humeau2013} the authors demonstrate an empirical plot of normalized root mean squared error against number of customers and show it decreases. 
This work aggregates  up to 782 homes.
In \cite{Silva2014} the authors show that mean absolute percentage error decreases with the number of customers and use it for examining electricity market trading performance. 
In \cite{Quilumba2015} the authors show that clustering can help improve the forecast accuracy, which follows the ideas proposed in this work.

This paper differs from prior work since we extend the aggregation to over 100,000 customers (100 MWh) and point out the crucial point that errors no longer improve beyond a critical load.
We then propose an additive load shape based consumption model based results from smart meter clustering analysis \cite{Kwac2014}.
This model is used to derive a benchmarking formula that describes the relationship between relative error and the aggregation size. 
The model is fit to experimental performance data of several forecasters to uncover aggregation relationships. 
To the best of our knowledge, this is the first paper modeling forecasting error scaling with aggregation size.
This work connects individual consumption models to aggregate forecasting and provides a simple mechanism to benchmark any forecasting algorithm as it is applied to varying levels of aggregation.

\section{Modeling Load Aggregation}
\label{section-Modeling-Load-Aggregation}

\begin{figure}[h] 
\centering
\subfigure[][]{
\includegraphics[scale=0.4]{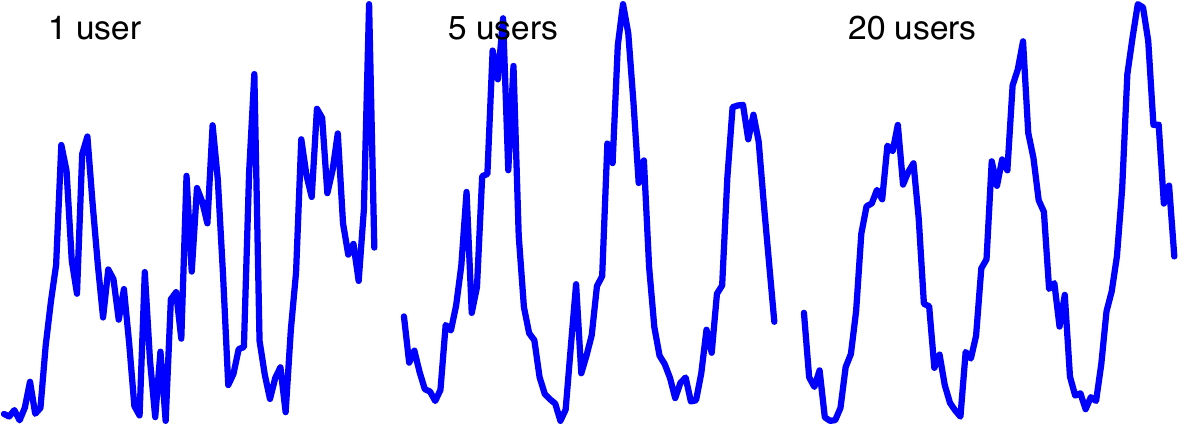}   
}
\subfigure[][]{
\includegraphics[scale=0.4]{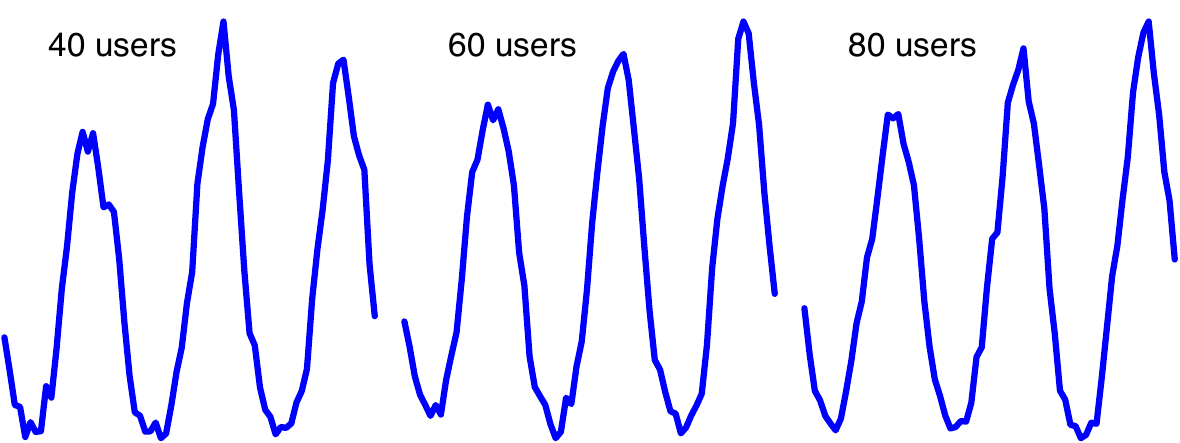}   
} 
\caption[]{
Hourly electricity consumption for various aggregation levels.  
Consumption pattern of a single customer generally has little structure to be exploited.  
Aggregating more and more customers `smoothes' the signal so that it can be more predictable.  
Aggregation level of 20 or more residential customers shows a predictable pattern.  
Note that plots are not in the same scale.
}
\label{fig:signal-at-aggregation-levels}
\end{figure}

Aggregation reduces the inherent variability in electricity consumption resulting in increasingly smooth load shapes. 
Figure~\ref{fig:signal-at-aggregation-levels} illustrates this effect where it is clear that the higher aggregation levels are easier to predict. 
An intuitive explanation is that the `law of large numbers' smoothes out the signal, therefore justifying why gigawatt level forecasting is very accurate.
Yet, it is less clear how to quantify the improvements in forecasting.
For example, it is assumed that more aggregation will generally improve forecast accuracy due to a $\frac{1}{\sqrt{N}}$ smoothing.

The main goal of this paper is to \emph{develop an scaling law for forecasting performance with respect to aggregation size} which fits experimental data. 
In particular, we propose that the mean load be used as a key metric in identifying how forecasting methods perform in very large population averages. 
We experimentally demonstrate the intuitive smoothing, but show that there is a limit with which aggregation no longer helps improve forecasting performance.
Finally, we propose a simple stochastic process model which describes the experimental Aggregate Error Curve.

\subsection{Forecast Accuracy Performance Metrics}
\label{section:One-Hour-Ahead-Univariate-Forecasting-for-Residential-Loads-On-Various-Levels-of-Aggregation}

The main performance metrics most commonly used in forecasting literature are Coefficient of Variation (CV) and Mean Absolute Percentage Error (MAPE). 
Recently, alternative metrics have been proposed \cite{Haben2014}, however we will not focus on these.
This work focuses mostly on CV, because it is amenable to theoretical analyst.
However, the proposed scaling law is fit to both CV and MAPE and they are shown to have identical behavior.

Coefficient of variation measures the ratio of the prediction error standard deviation to the signal mean.  
Consider two time series $x(t)$ and its forecast $\hat{x}(t)$ for $t=\{1,...,T\}$. 
The empirical CV measures the difference between these time series and is computed as 

\begin{align}
CV(x, \hat{x}) = 100 \frac{\sqrt{ \frac{1}{T} \Sigma_{t=1}^{T} (x(t) - \hat{x}(t) )^{2} } } { \frac{1}{T} \Sigma_{t=1}^{T} x(t)} \%. 
\end{align} 
Likewise the mean absolute percentage error (MAPE)  is defined as
\begin{align}
MAPE(x, \hat{x}) = \frac{100}{T} \sum_{i=1}^{T} \left| \frac{ x(t)  - \hat{x}(t)   } { x(t)  } \right| \%.
\end{align}

CV and MAPE are relative error metrics traditionally reported in the literature.
It is assumed they allow comparison of performance in different datasets because they are relative metrics.
However as we show in this work, the relative error depends significantly on the level of load aggregation.
\subsection{Forecasting Scaling Laws}
\label{subsection:Forecasting-Scaling-Laws}

Consider a set of $N$ customers with consumption given by a time-series $x_n(t)$. 
The mean consumption for each customer is $W_n = \frac{1}{T}\sum_{t=1}^T x_n(t).$
We select a subset $A \subseteq \{1, 2, \hdots, N \}$ of customers and form an aggregation consumption 
\begin{align}
  x_{A}(t) = \sum_{n \in A} x_n(t), \label{eq:aggregate-definition}
\end{align}
\noindent with average consumption
\begin{equation} 
W_A = \sum_{n \in A} W_n.
\label{eq:avg_consumption}
\end{equation} 
In load forecasting, we build a predictor for  $x_{A}(t)$ that outputs  the predicted sequence $\hat{x}_{A}(t)$  and evaluate  CV$(x_A, \hat{x}_A)$. 
Suppose such forecaster is built for every such group of customers.  
Then, under certain conditions on the behavior of $x(t)$ and the forecasting model, we have the following: 
\begin{thm} 
\label{expectation-upper-bound-theorem} 
Consider all sets $A$ of consumers $x_{A}(t)$  with mean consumption $W_{A} = W$. 
The average coefficient of variation at the $W$ level of aggregation is given by
\begin{align}
CV(W) &= \textbf{E}[ \text{CV}(x_A, \hat{x}_{A})| W_A = W] \\ 
            &=\sqrt{ \frac{\alpha_0}{W} + \alpha_1 } \label{eq-mean-CV-W} 
\end{align} 
 for constants  $\alpha_0$ and $\alpha_1$, where the expectation is taken over all sets $A$ of mean consumption $W$.
\end{thm}
\begin{proof}  
See \ref{appendix-analytical-model-of-aggregation}
\end{proof}

Theorem \ref{expectation-upper-bound-theorem} gives a set of sufficient conditions under which the coefficient of variation will converge to \eqref{eq-mean-CV-W}.
The conditions are quite standard, and produce an intuitive scaling law.
However, as will be shown, this law fits the experimental aggregation with some error.
We propose a modification of the population average \emph{CV scales as a function of $W$} following  
\begin{align}
CV(W) =  \sqrt{ \frac{\alpha_{0}}{W^p} + \alpha_1 }  \label{eq:CV-with-load}
\end{align}
to fit the AEC's produced for many of the forecasting horizons.
The parameter $p$ is used to provide flexibility in fitting experimental aggregation error curves.
Note that an \textit{ideal aggregation} occurs when $p=1$ and $\alpha_1=0$.
In this ideal case, we have the intuitive $\frac{1}{\sqrt{W}}$, improvement that is naively assumed to be true.
Assuming $p=1$, the proposed scaling law segments the forecasting problem into two regimes: 
\begin{enumerate}
\item \emph{Scaling}: When ${\alpha_{0}/\sqrt{W}} \gg\alpha_1$, relative error improves considerably due to aggregation. 
Equation \eqref{eq:CV-with-load} can be approximated as
$CV(W) \approx \sqrt{\frac{{\alpha_0}}{{W}}}.$
\item \emph{Saturation}: 
When ${\alpha_{0}}/W\ll\alpha_1$, there is no improvement in forecasting from aggregation and $\text{CV}(W) \approx \sqrt{\alpha_{1}}$.
\end{enumerate}

In addition to CV, we extend our proposed model for MAPE as well.
We fit the population average \emph{MAPE scales as a function of $W_A$} according to  
\begin{align}
\textrm{MAPE}(W) &= \textbf{E}[\textrm{MAPE}(x_A, \hat{x}_A)\left| W_A = W \right.]\nonumber\\ 
&=  \sqrt{ \frac{\beta_{0}}{W^p} + \beta_1 }.  \label{eq:MAPE-with-load}
\end{align}
 
For all of the experiments, both the CV and MAPE aggregation error curves were generated, but the CV based AEC's are shown in the figures, since they correspond to the theoretical analysis in \ref{appendix-analytical-model-of-aggregation}. 

%
%
%
%
%

\section{Experiment Setup}
\label{section-Experiment-Setup}

\subsection{Description of Data}
\label{subsection-Description-of-Data}

\begin{figure}[h]
\centering
\subfigure[][]{
\label{fig:data-description-a}
\includegraphics[width=0.215\textwidth, height=0.215\textwidth]{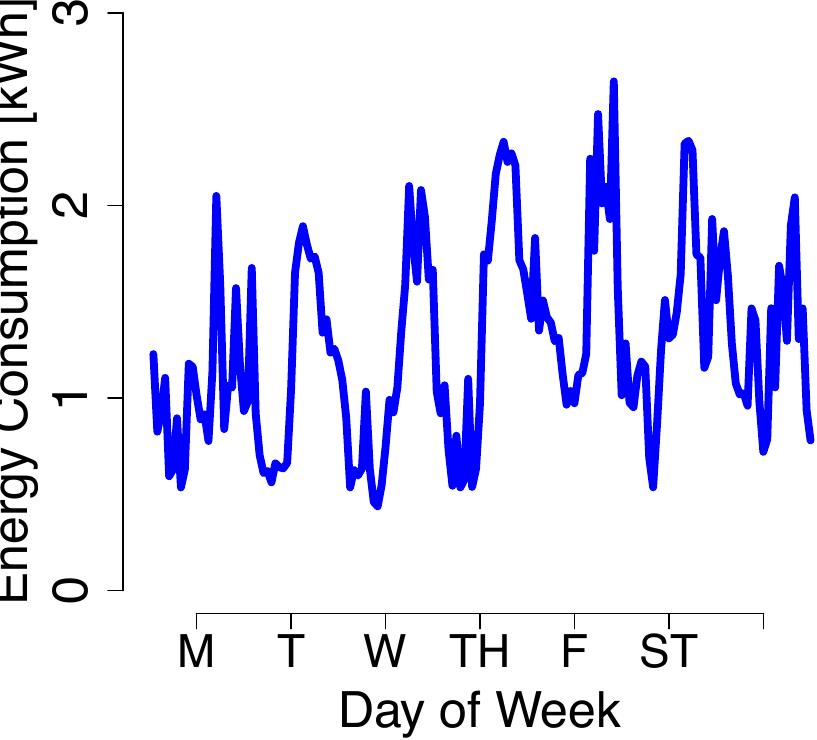}     }
\hspace{8pt}
\subfigure[][]{
\label{fig:data-description-b}
\includegraphics[width=0.215\textwidth, height=0.215\textwidth]{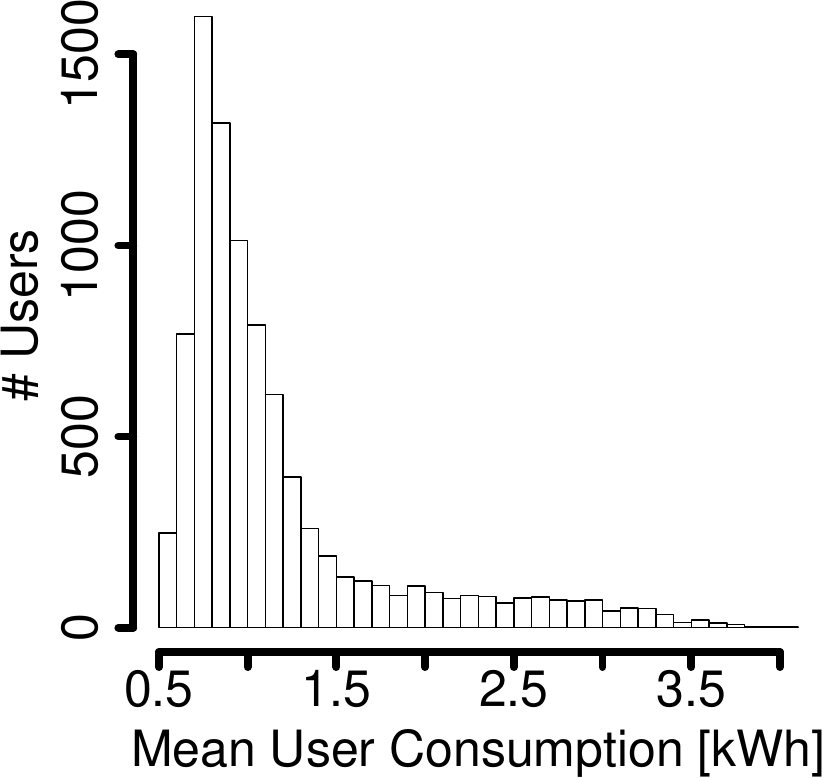}    
} 
\subfigure[][]{
\label{fig:data-description-c}
\includegraphics[width=0.215\textwidth, height=0.215\textwidth]{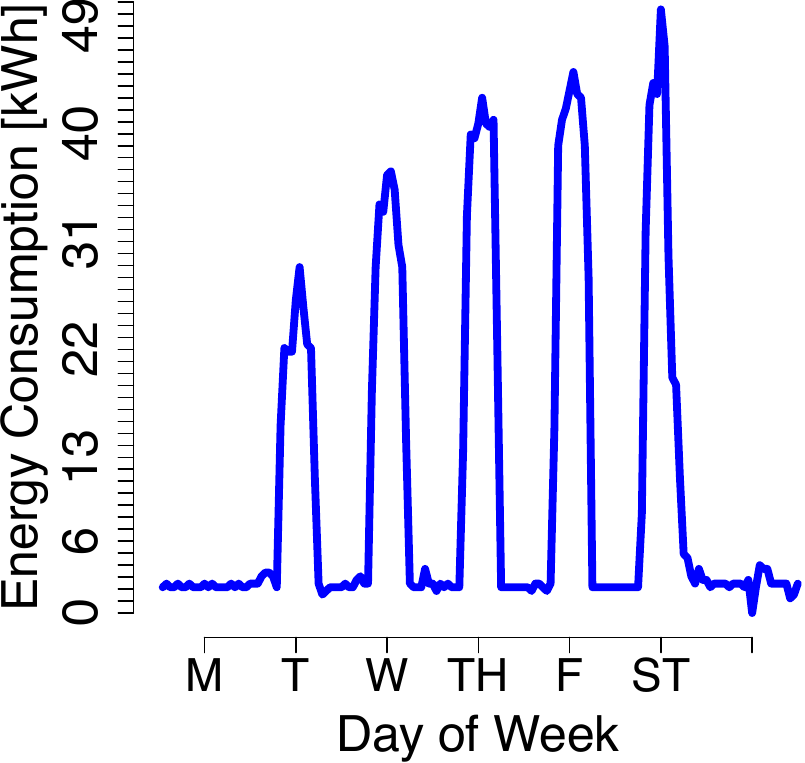} }
\hspace{8pt}
\subfigure[][]{
\label{fig:data-description-d}
\includegraphics[width=0.215\textwidth, height=0.215\textwidth]{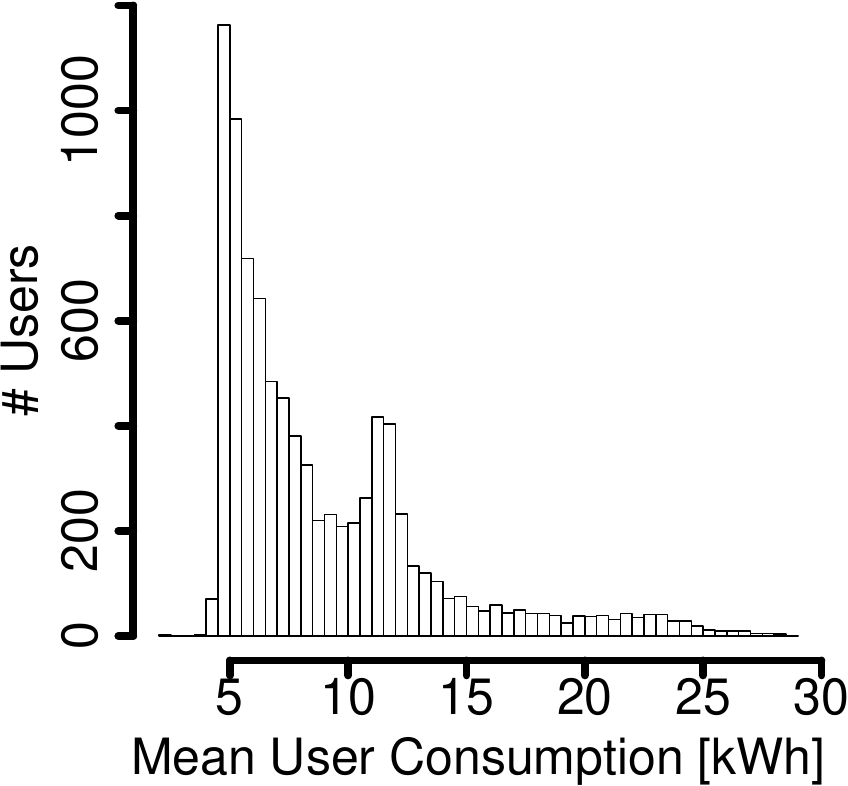} 
}
\caption[Residential and SMB data statistics.]{
7 days of consumption for single \subref{fig:data-description-a} residential customer and \subref{fig:data-description-d} SMB customer.
\subref{fig:data-description-b} Histogram of mean load for all \subref{fig:data-description-b} residential and \subref{fig:data-description-c} SMB customers.
}
\label{fig:data-description}%
\end{figure}

The data used for our study is provided by Pacific Gas and Electric Company (PG\&E).  
With a customer base of residential (RES) and small to medium businesses (SMB).
Residential customer data comes from most of northern California as 1 hour intervals meter reads which are used for billing.
The entire dataset represents over $180$ thousand users of year long consumption from 2010-08-01 from 2011-07-31.  
The data represents $408$ zip codes around California out of a total of $2,597$ possible zip codes.  
The SMB data comprises of $150$ thousand consumer profiles.  
The data set represents a full year from 2010-08-01 to 2011-07-31, sampled every $15$ minutes. 
The data represents a wide variety of commercial applications, in many ZIP codes and climate profiles.  
For this paper, the data is temporally aggregated to represent 1 hour interval data like the residential data.
   
The mean consumption of the data is of importance for this work.
Figure \ref{fig:data-description-a} shows a typical one week time series of a residential customer, while Figure \ref{fig:data-description-b} shows the consumption of each residential customer averaged over an entire year.
The overall mean consumption is $1.05$ kWh.  
Although there is some variation, the maximum mean consumption is less than $4$ kWh.  
The SMB data differs from the residential data in two ways:
(1) The mean consumption is generally much larger than a residential customer;
(2) the profiles are less variable, in terms of intra-day variation.  
The mean consumption for the SMB data is $8.94$ kWh, which is close to 9 times that of a residential customer.
Figure \ref{fig:data-description-c} shows the consumption profile of a randomly chosen customer and Figure \ref{fig:data-description-d} shows the histogram of yearly means for all users.

\subsection{Generating Aggregate Consumption}
\label{subsection-Generating-Aggregate-Consumption}

The following procedure is used to generate an Aggregation Error Curve for a given population and forecasting procedure.
  
\begin{enumerate}
\item A set of aggregate profiles are generated using \eqref{eq:aggregate-definition}, by randomly sampling the population of profiles $A$.
For a fixed array of aggregation sizes, $\mc{N} = \{N_1 \hdots N_{max}\}$, we choose a size, $N$ then sample without replacement a subset $A$, where $|A| = N$.
This is done a fixed number of times $M$. 
Therefore for any $N$, and $m \in \{1, \hdots, M \}$ a single aggregation $x_{A}$ is generated.
For each aggregate $A$, the mean consumption is computed according to \eqref{eq:avg_consumption}.
\item A forecasting procedure is used to generate $\hat{x}_{A}$ is computed for each aggregate signal and each time $t$.
\item Error metrics are computed with either $\text{CV}(x, \hat{x})$ or $\text{MAPE}(x, \hat{x})$.
\item For each $N$ and $m$ the tuple $(W_{A}, CV(x, \hat{x}) )$ is recorded.
\end{enumerate}

Residential aggregate consumption time series were generated by forming groups of randomly selected customers.
Fifty six group sizes were chosen ranging from one to 100,000 customers, the values of aggregation sizes are given in \ref{Appendix-Aggregation Sizes}.
Fifty random groups for each size were generated by uniformly selecting customers. 
The mean hourly consumption of these groups ranged from 1kWh to 100MWh. 
The largest mean hourly consumption for each size ranged from  3 kWh to 180 MWh. 
For each generated aggregate load, we generate a weighted average temperature time series for each zip code used in generating an aggregate.
   
SMB aggregate consumption time series are generated in a similar way. 
Forty three group sizes were selected ranging from one to 50,000 customers. 
The hourly group mean consumption ranges between 10kWh to 400MWh. 
The group with largest consumption had a $670$ MWh average hourly load. 


\subsection{Forecasting Models}
\label{section:Forecasting Models}

\begin{table}[h]
\centering
\caption{Models used in analysis} 
\label{tab:model-comparison}
\begin{tabular}{@{}ccl@{}}
\toprule
Model & \phantom{abc} & \    Description \\
\cmidrule{1-1} \cmidrule{3-3}
$\mc{M}_1$  &&  $\text{ SARMA}(1, 0){\times}(1, 0)_{24} $    \\ 
$\mc{M}_2$  &&  $\text{ SARMA}(2, 0){\times}(1, 0)_{24}$     \\ 
$\mc{M}_3$  &&  $\text{ SARMA}(3, 0){\times}(1, 0)_{24}$     \\ 
$\mc{M}_4$  &&   \   SVR ~~- Radial Basis Function               \\ 
$\mc{M}_5$  &&   \   FFNN  - Logistic Activation Function        \\ 
$\mc{M}_6$  &&   \   Daily Total (SARMA) + Shape forecast     \\ 
\bottomrule
\end{tabular}
\end{table}
The proposed scaling laws are studied using three commonly used methods for short term load forecasting: Seasonal Auto Regressive Moving Average (SARMA), Support Vector Regression (SVR) and Feed Forward Neural Networks (FFNN).  
For models $\mc{M}_1 \hdots \mc{M}_5$, a one hour and multiple hour ahead forecasting problem is studied.
Model $\mc{M}_6$ is used for the full day ahead forecasting experiment.
Each model and their training/testing procedures are described in detail in the \ref{Appendix-Forecasting-Models}.

\section{Experimental Results}
\label{section:Experimental-Results}
\subsection{Empirical CV and Aggregation Level} 
\label{sub-section:Empirical-CV-with-Aggregation-Level}

\begin{figure}[h]
\centering
\includegraphics[width=0.5\textwidth, height=0.4\textwidth]{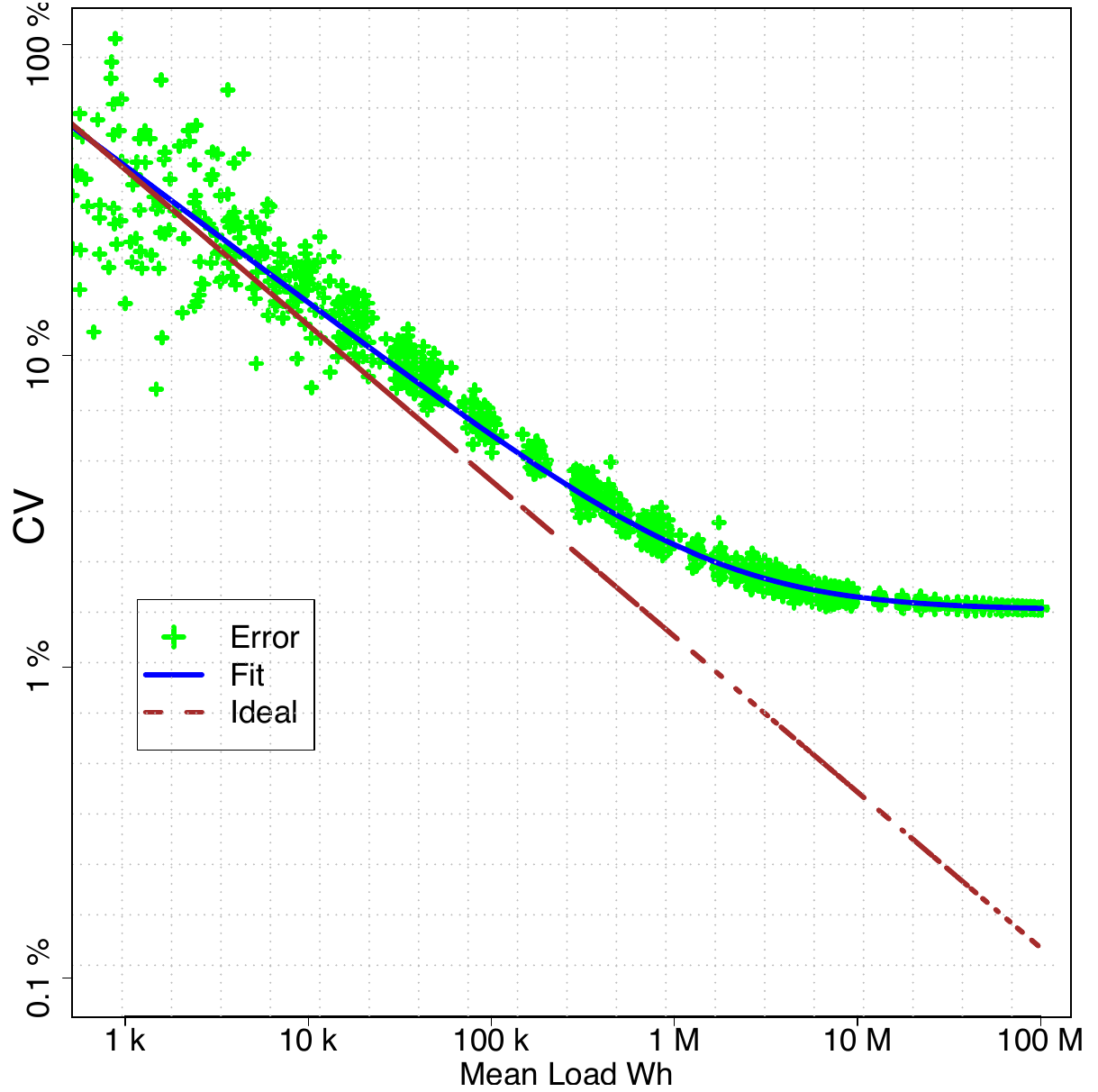}
\caption{ 
$(W_A, \text{CV}(x_{A}, \hat{x}_{A}))$ is shown in green markers for model $\mc{M}_1$.
Best fit (solid line) has $p=0.89$.  
Dashed line indicated error scaling with ideal aggregation effect with no irreducible error ($p=1$).  
The model leads to a critical load, $W^{\star}=2179$ kWh, and irreducible error, $\sqrt{\alpha_1}=2.1$.
}
\label{fig:unv-M1-log-log}   
\end{figure}
     
First we investigate the performance of model $\mc{M}_1$ in the one hour ahead prediction task. 
For each time-series, the corresponding mean load $W_{A}$,  forecast $\hat{x}_{A}$ and performance $\text{CV}(x_{A}, \hat{x}_{A})$ were computed. 
The Aggregation Error Curve is the plot of the pairs $(W_A, \text{CV}(x_{A}, \hat{x}_{A}))$ for all generated groups $A$ as shown in Figure~\ref{fig:unv-M1-log-log}. 
The scaling law in \eqref{eq:CV-with-load} was fit to this data using a non-linear least-squares procedure detailed in \ref{appendix-parameter-fit-NLLS}.
The fit (solid line) and the \emph{ideal aggregation} scaling laws are displayed in Figure \ref{fig:unv-M1-log-log}. 
Fit parameters are shown in Table~\ref{tab:cv-mape-model-comparison} for model $\mc{M}_1$ and others.

Figure \ref{fig:unv-M1-log-log} shows visual verification that the scaling law can be decomposed into scaling and saturation regimes.
The transition point between regimes is defined as the load aggregation level $W^{\star}$ where the regime approximations are equal.  
The critical load $W^{\star}$ is the positive solution $W$ to
\begin{align}
\sqrt{\frac{\beta_0}{W^p}} = \sqrt{\beta_1}. 
\end{align}
The critical load for model $\mc{M}_1$ is $2.2$ MWh.
The \emph{scaling} regime extends from $1$ kWh to $2.2$ MWh aggregate loads, and the \emph{saturation} regime extends from $2.2$ MWh to $100$ MWh.   

Note that any intermediate sized aggregate of loads will fall somewhere in the scaling regime of the AEC since the scaling regime is sensitive to aggregation size. 
Because of this, understanding how aggregation leads to forecasting improvement is important in designing applications which require small to medium size aggregations of customers.
  
\subsubsection{Comparison of Different Models}
\label{sub-section:Comparison-of-Different-Models} 

\begin{table}[h]
\centering
\caption{Scaling law fit for CV} 
\label{tab:cv-mape-model-comparison}
\begin{tabular}{@{}ccccccccc@{}}
\toprule
             & \multicolumn{4}{c}{CV}                                                                                                                                        & \multicolumn{4}{c}{MAPE}  \\
$\mc{M}$ &   p                                          & $\sqrt{\alpha_0}$  & $\sqrt{\alpha_1}$   & $W^{\star}$    &    p                                       & $\sqrt{\beta_0}$                   & $\sqrt{\beta_1}$                          & $W^{\star}$     \\  
\cmidrule{1-9}
$1$      &  0.89 {\scriptsize(0.83 0.96)} & 53.8 {\scriptsize(50.0 60.7)} & 2.13 {\scriptsize(2.13 2.13)} & 2179   & 0.88 {\scriptsize(0.83 0.95)} & 45.4 {\scriptsize(50.0 60.7)} & 1.52 {\scriptsize(1.486   1.53)}   & 2250   					    \\ 
$2$      &  0.92 {\scriptsize(0.87 0.98)} & 53.5 {\scriptsize(48.5 58.8)} & 1.25 {\scriptsize(1.20 1.31)} & 8925   & 0.96 {\scriptsize(0.92 1.05)} & 43.6 {\scriptsize(38.2 51.9)} & 0.89 {\scriptsize (0.858  0.970)} & 8621   					    \\ 
$3$      &  0.91 {\scriptsize(0.87 0.96)} & 54.4 {\scriptsize(45.9 57.5)} & 1.19 {\scriptsize(1.14 1.23)} & 11615 & 0.95 {\scriptsize(0.90 1.02)} & 42.1 {\scriptsize(39.6 50.6)} & 0.81 {\scriptsize (0.816  0.820)} & 16856  					    \\ 
$4$      &  0.92 {\scriptsize(0.81 1.00)} & 52.9 {\scriptsize(48.6 53.8)} & 1.96 {\scriptsize(1.95 2.06)} & 6089   & 1.07 {\scriptsize(0.99 1.15)} & 56.7 {\scriptsize(41.1 55.4)} & 1.21 {\scriptsize (1.081  1.913)} & 23127 					    \\ 
$5$      &  0.92 {\scriptsize(0.85 0.99)} & 55.8 {\scriptsize(49.3 59.9)} & 1.33 {\scriptsize(1.13 1.53)} & 72218 & 0.94 {\scriptsize(0.83 1.05)} & 45.3 {\scriptsize(37.0 53.6)} & 1.42 {\scriptsize (1.120  1.853)} & 1539   				           \\  
\bottomrule
\end{tabular}
\end{table}
  
In this section we validate that the scaling law holds for the models in Section \ref{section:Forecasting Models}.
The scaling law parameters also provide a way to compare the performance of these different models. 
For the models in Table \ref{tab:model-comparison} the resulting scaling law fits for MAPE and CV are given in Table~\ref{tab:cv-mape-model-comparison}.
 
The relative variation of the parameters $\sqrt{\alpha_0}$ and $\sqrt{\beta_0}$ are quiet small between different models.
In contrast, the irreducible errors $\sqrt{\alpha_1}$ and $\sqrt{\beta_1}$ are much larger different between the models.
As is shown in \ref{appendix-analytical-model-of-aggregation}, the model implies that $\sqrt{\alpha_0}$ and $\sqrt{\beta_0}$ should be identical since they are associated with independent errors.
On the other hand $\sqrt{\alpha_1}$ and $\sqrt{\beta_1}$ capture the true forecasting performance.
Combining these two important observations leads to identifying the irreducible errors as a fundamental performance metric for model comparison. 

Using this metric, we see that with sufficient training, the SVR and FFNN models perform quite well.
These models show irreducible CV errors of  $1.961\%$ and $1.338 \%$, respectively and MAPE errors of $1.210 \%$ and $1.423 \%$.
We should note however, that these models take considerably more training since a number of running parameters are fit in the validation step prior to a single test sample is evaluated.
Additionally, over a large sample population the linear models outperform these more computationally intensive models.

The critical load value $W^*$ can be compared between different models. 
It can be seen to depend almost exclusively on the irreducible error since the reducible errors are close to each other. 
This observation leads to the conclusion that \textit{forecasters with low irreducible error benefit more from aggregation}. 
For example, Model $\mc{M}_3$ has critical load of $16$ MWh and it's saturation regime is in the far right of the AEC.

Table \ref{tab:cv-mape-model-comparison} shows the fitted values as well as $95\%$ confidence intervals which are computed by bootstrap resampling the experimental points \cite{Efron1982}.
Two key points are shown by the confidence interval:
\begin{enumerate}
\item The confidence intervals of $\sqrt{\alpha_0}$ and $\sqrt{\beta_0}$ are quite wide and intersect in the intervals $\sqrt{\alpha_0} \in [50.0,~53.8 ]$ and $\sqrt{\beta_0} \in [50.0,~50.6]$.  
This means that for example, a null hypothesis, halfway between the ranges, of $\sqrt{\alpha_0} = 52.4$ and $\sqrt{\beta_0} = 50.3$ identical over all models would not be rejected.
This validates the analytical model relating this reducible term to a consumption profile independent parameter.
\textit{A somewhat controversial conclusion from this is that most forecasting methods modeling data at small aggregation levels are merely overfitting random noise.}
\item The value's of $p$ for each fit are interesting since for an individual dataset, setting $p=1$ leads to an unseemly fit.
However, for half of the models, the $95 \%$ confidence interval contains the value $p=1$.
Therefore we conclude, that for a given dataset, generating an accurate fit of the aggregation-error curve requires a $p \neq 1$.
However, we should keep in mind that there is no model basis for this parameter.
\end{enumerate} 

\subsection{Linear Scale Observations}
\label{Linear Scale Observations}

\begin{figure}[h]    
\centering
\subfigure[][]{
\includegraphics[width=0.40\textwidth, height=0.38\textwidth]{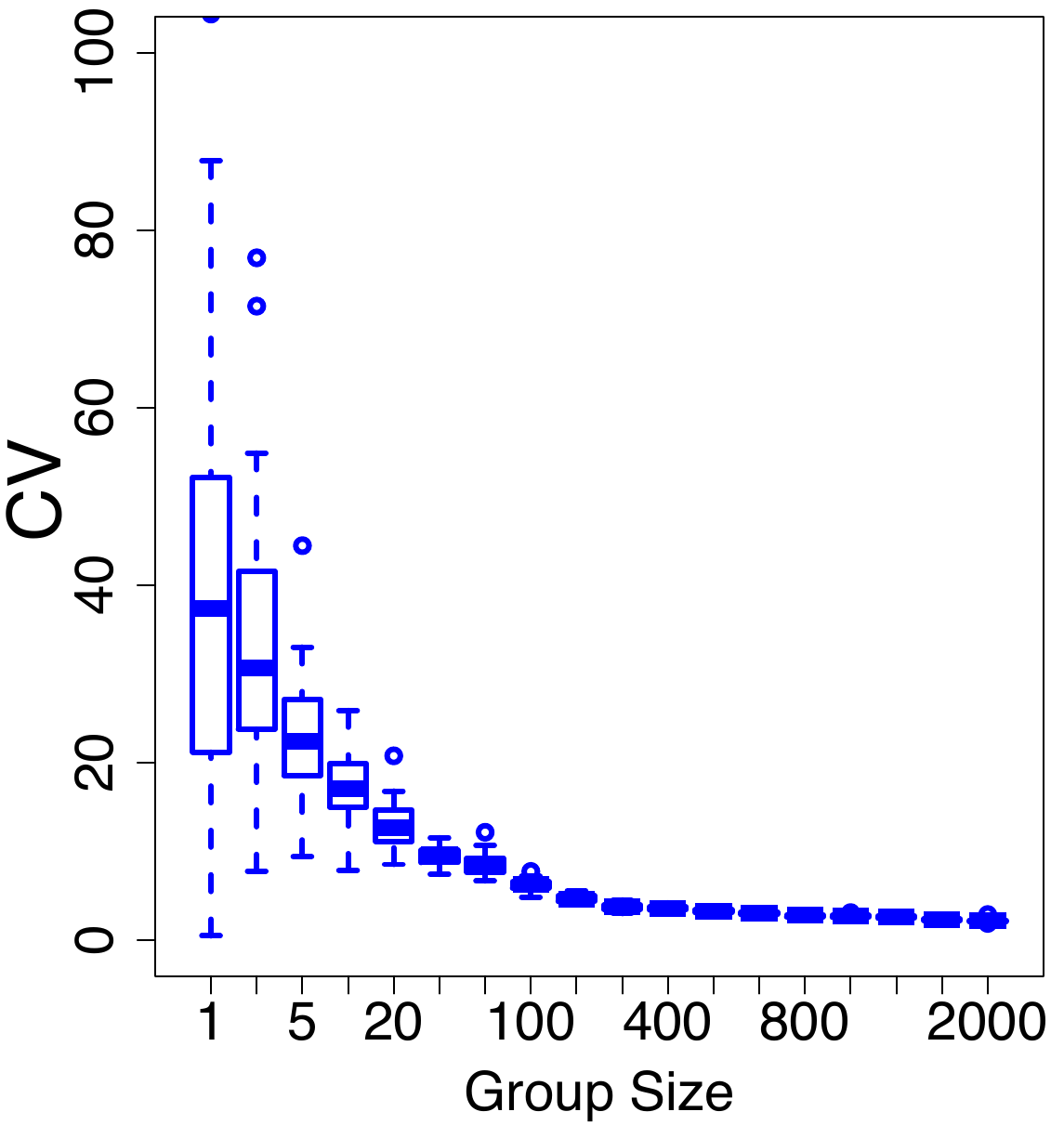}     
\label{fig:lin-scale-obs}
}
\subfigure[][]{
\includegraphics[width=0.4\textwidth, height=0.38\textwidth]{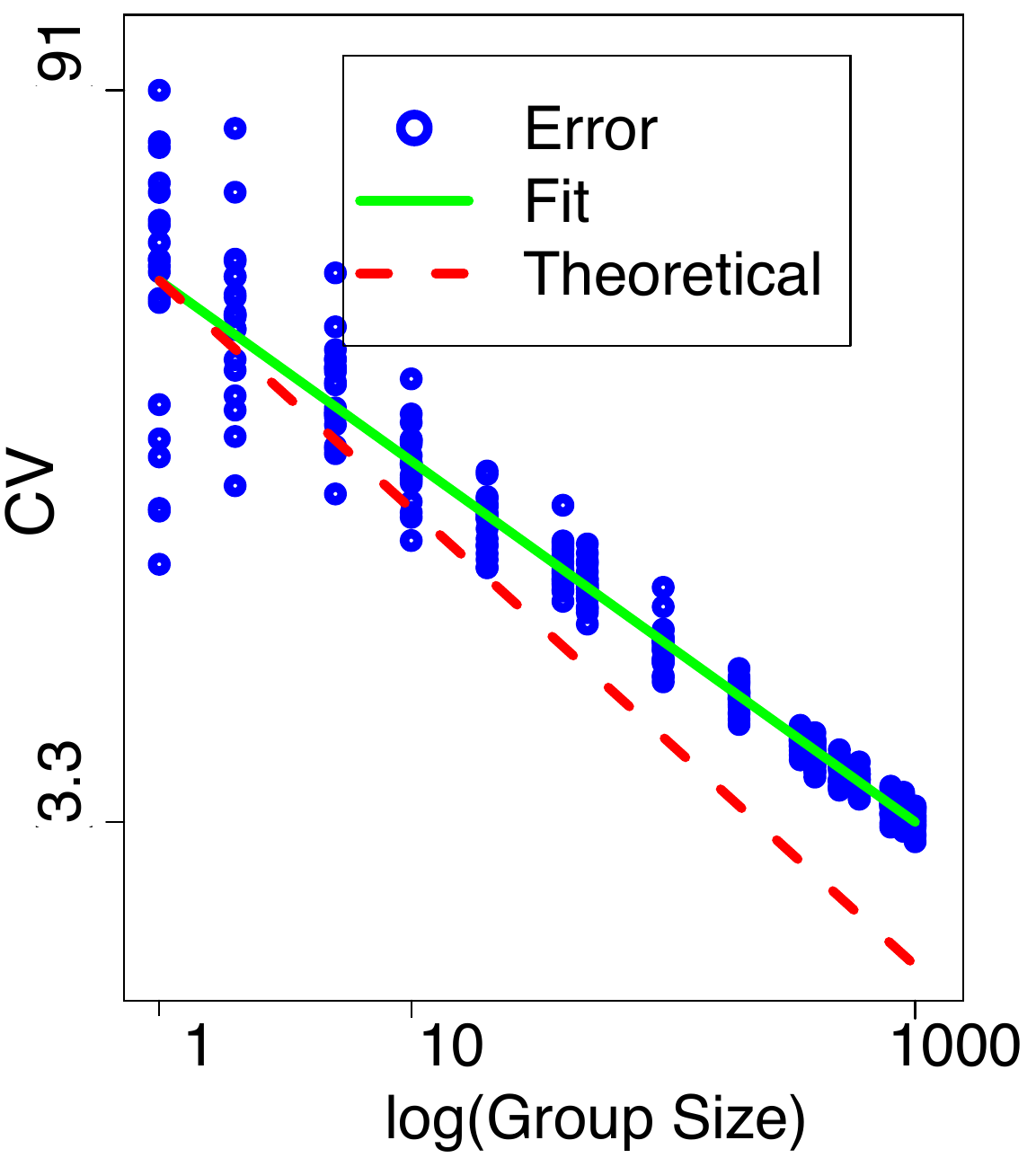}
\label{fig:log-scale-obs}
}   
\caption[A set of four subfigures.] {
CV boxplot of randomly generated groups of N customers in linear \subref{fig:lin-scale-obs} and log scale \subref{fig:log-scale-obs}.
}
\end{figure}  

Here we investigate the effect of aggregation in a linear scale and small sample sizes to show how the proposed model deviates from common understanding of aggregation smoothing.
For example, analysis of $2000$ customers was presented in \cite{Sevlian2013} and with $200$ customers in \cite{Silva2014}. 
Such limited group sizes are unable to identify the various regimes present in the scaling law, thus an incomplete understanding.

Most studies that have shown similar results to this work have relied on a few data points and have shown that relative error decreases as the aggregation size has increased.
All prior work have shown a linear scale decrease which has fit an intuition of the ``law of large numbers".
It should be stressed that the analysis in the log scale, such as in Figure \ref{fig:unv-M1-log-log}, provides a different intuition than a linear scale analysis.
For example, the log scale analysis visually highlights the existence of a critical load and irreducible error.
This is not noticeable in linear scale analysis with small sample sizes. 
The same linear CV plot is replotted in a log scale Figure \ref{fig:log-scale-obs}.
Notice that even though a rough ``diminishing returns" is assumed in Figure \ref{fig:lin-scale-obs}, it is not seen in Figure \ref{fig:log-scale-obs}.
The errors are still decreasing at a constant rate, which is noticeable only in a log-log plot and only for larger aggregation levels.

\subsection{Day Ahead and Multiple Hour Forecasting}
\label{subsection-multiiple-hour-ahead-and-day-ahead-forecasting}

Here we present forecasting Aggregation Error Curves for both multi-hour horizon and complete day ahead forecasting problem.
We show that that the model of aggregation is consistent regardless of forecasting horizon if the same method is used over all time series.

\begin{figure}[h]  
\centering
\subfigure[][]{
\includegraphics[width=0.45\textwidth, height=0.4\textwidth]{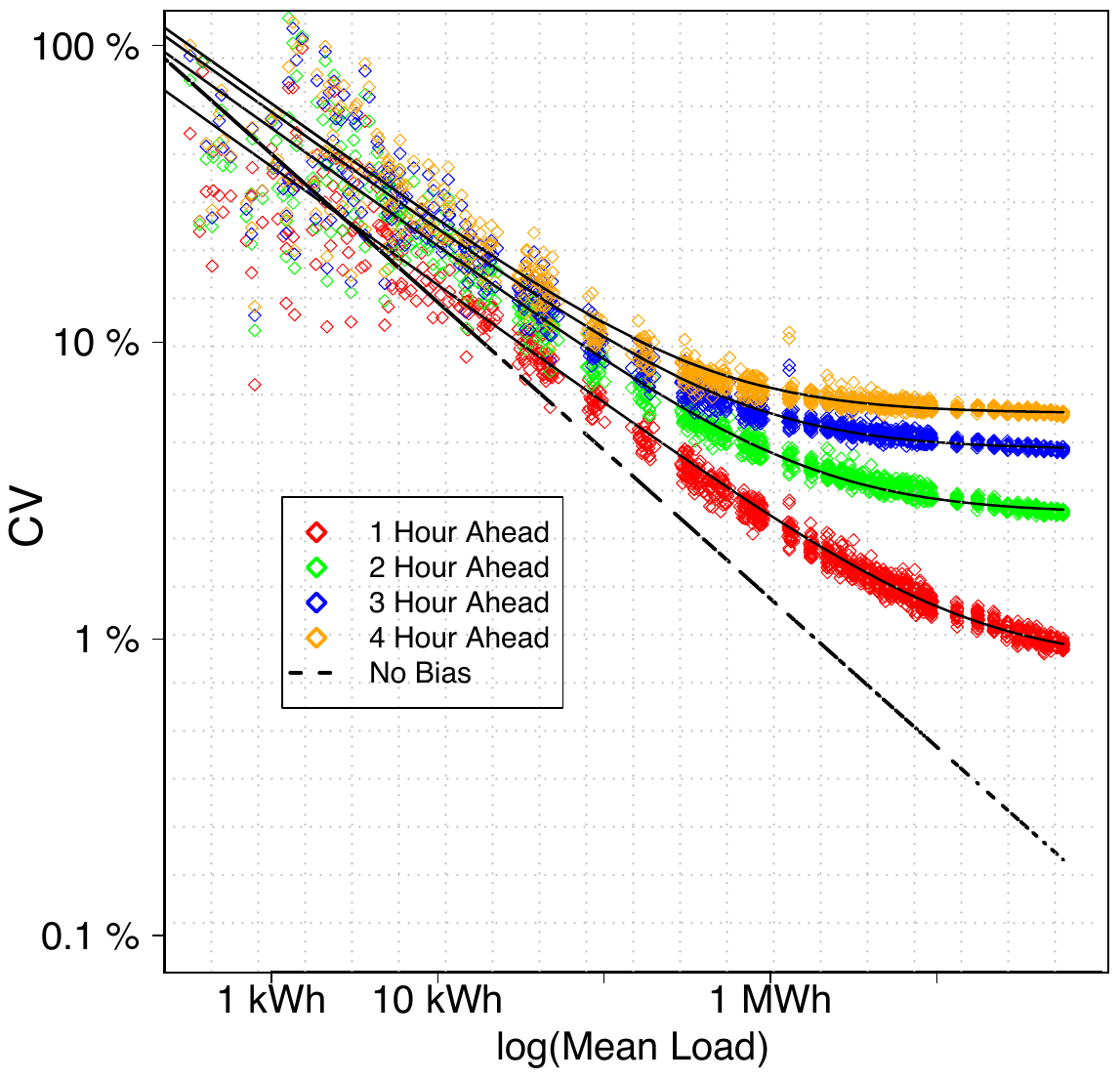}
\label{fig:n-hour-ahead-CV-res-compare}
}
\subfigure[][] { 
\includegraphics[width=0.45\textwidth, height=0.4\textwidth]{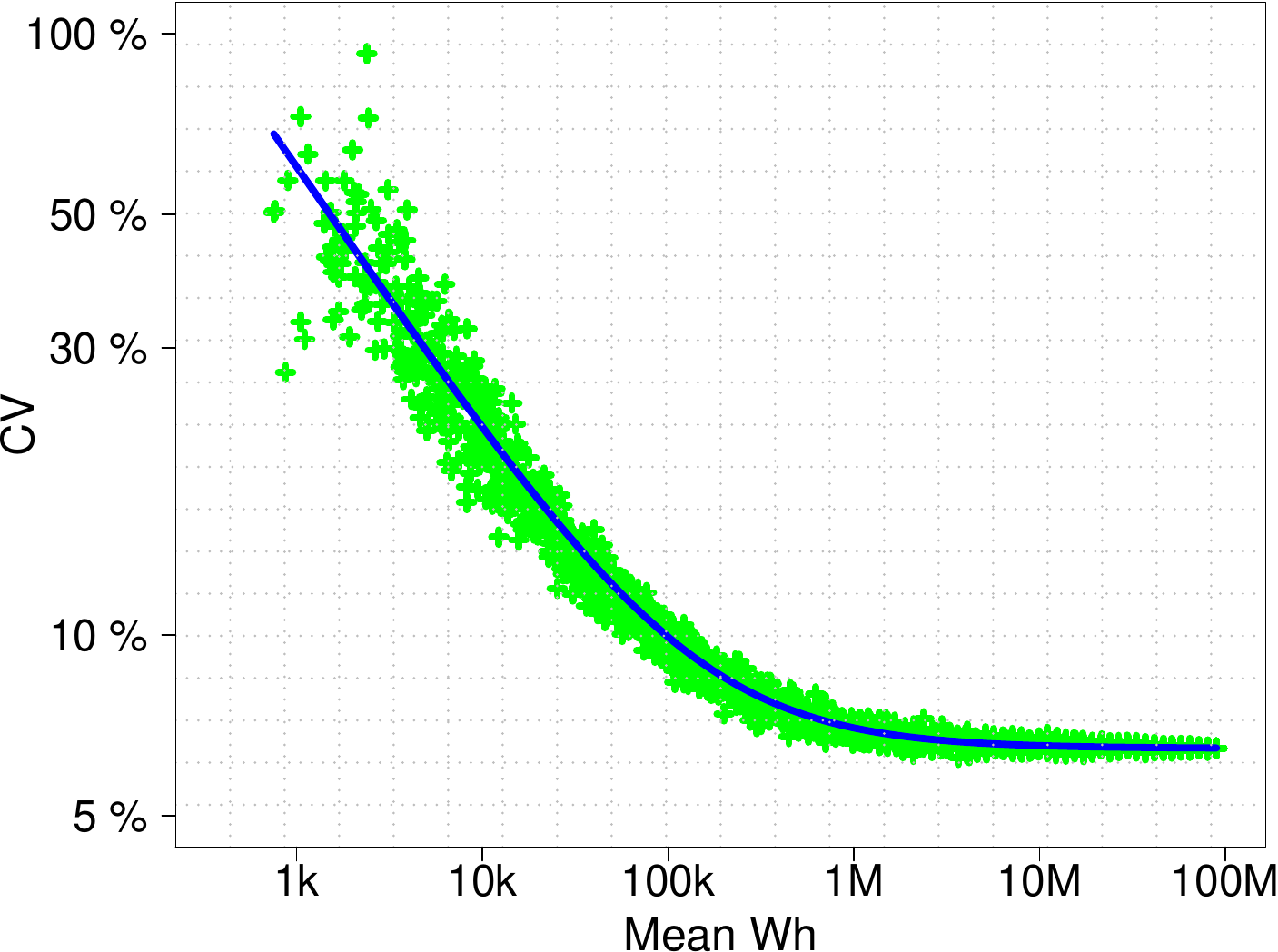}
\label{fig:day-ahead-aggregation-error-fit}
}
\caption[]{
\subref{fig:n-hour-ahead-CV-res-compare} AEC for multiple hours ahead using model $\mc{M}_3$.
\subref{fig:day-ahead-aggregation-error-fit} AEC for day ahead forecaster $\mc{M}_6$.
}  
\end{figure}

\subsubsection{Multiple Hour Forecasting}

Figure \ref{fig:n-hour-ahead-CV-res-compare} shows the model $\mathcal{M}_3$ used for various hour ahead forecasting problems.
The scaling law is fit to the data, with fit parameters given in Table \ref{table:N-hour-ahead-model-fitting-m3}.
It is clear that the irreducible error increases with forecasting horizon, reducing the benefit from aggregation. 
It indicates that for more complex tasks such as day ahead forecasting, the forecasters need to be designed carefully to achieve low irreducible error.

\begin{table}[h]
\centering
\caption{Scaling law fit for multiple horizons} 
\label{table:N-hour-ahead-model-fitting-m3}
\begin{tabular}{@{}crrcrr@{}}\toprule
Horizon & \multicolumn{2}{c}{CV} & \phantom{abc}& \multicolumn{2}{c}{MAPE}\\
\cmidrule{2-3} \cmidrule{5-6}
 (hours ahead) & $\sqrt{\alpha_{0}}$   & $\sqrt{\alpha_{1}}$ && $\sqrt{\beta_{0}}$  & $\sqrt{\beta_{1}}$ \\ \midrule
1  &  72.5  &  1.28   &&   56.3    &   1.06   \\ 
2   & 85.2  &  5.30   &&   75.1    &   3.26   \\ 
3  &  50.2  &  8.92   &&   86.6    &    7.35  \\ 
4  &  68.3  & 10.94   &&   78.6    &   8.11  \\ 
\bottomrule
\end{tabular}
\end{table}

\subsubsection{Day Ahead Forecasting}
\label{section:Day-Ahead-Forecasting}
  
The aggregation error curve for randomly generated loads using model $\mc{M}_6$ is shown in Figure \ref{fig:day-ahead-aggregation-error-fit}.
Using our proposed model, we have the following approximate model 

\begin{align}  
\textrm{CV}(W) =  \sqrt{ \frac{3562 }{W} + 41.9} \label{eq:day_ahead_forecast}.
\end{align}

The fit exponent $p = 1.01$, with the $95 \%$ confidence interval containing $p = 1$ while the reducible error is $\sqrt{3562} = 59.6$.
This leads to an irreducible error of $6.47 \%$ for a full day ahead forecaster and a critical load of $W^{\star} = 85$ kWh, or around 80 homes.  
This value may be misleading, since from Figure \ref{fig:day-ahead-aggregation-error-fit}, it appears as though forecast performance flattens our after $1$ MWh of load corresponding to $1000$ homes.
Here we can note that many forecasting methods shown in the literature clearly outperform the $6.47 \%$ value since they are finely tuned to a small dataset. 
When the static forecasting model is applied to a large number of randomly generated time series, the average performance is degradated.

\subsection{Small and Medium Business (SMB) Data Analysis}
\label{section:SMB-Data-Analysis}

The analysis in Section \ref{sub-section:Empirical-CV-with-Aggregation-Level} is extended to the SMB dataset by computing the CV scaling law for model $\mc{M}_3$. 
The obtained parameters are $\sqrt{\alpha_0} = 46.67$,  $\sqrt{\alpha_1} = 0.92$ and $p=0.82$. 
The scaling law can be compared to the residential dataset scaling law for the same model. 
The AEC for model $\mc{M}_3$ are shown in Figure \ref{fig:n-hour-ahead-MAPE-and-smb-res-compare-b} and are very close to those obtained for the same model used in the residential dataset.
Notice the scaling law is quite consistent despite the mean loads of residential and SMB data differing by an order of magnitude. 

This observation validates the choice of kWh average to drive the scaling rather than the number of customers. 
An intuitive interpretation of similarity in the scaling law is that every building (residential or SMB) consumes electricity as a series of tasks of similar average sizes.  
Larger buildings can be thought of as an aggregation of smaller buildings, so the number of tasks scale linearly and so does average consumption. 
This leads to kWh providing the proper scaling for forecasting. Finally, we note that for the SMB data the critical load is close to $10$ MWh. 
Forecasting studies in commercial buildings (mean loads $100$ kWh to $1$ MWh) need to consider the improvements due to aggregation when compared to each other.
  
%
%

\subsection{Subpopulation Comparison}
\label{Subpopulation Comparison}

\begin{figure}[h]
\hspace{-8mm}
\subfigure[][] {   
\label{fig:n-hour-ahead-MAPE-and-smb-res-compare-b}
\includegraphics[scale=0.24]{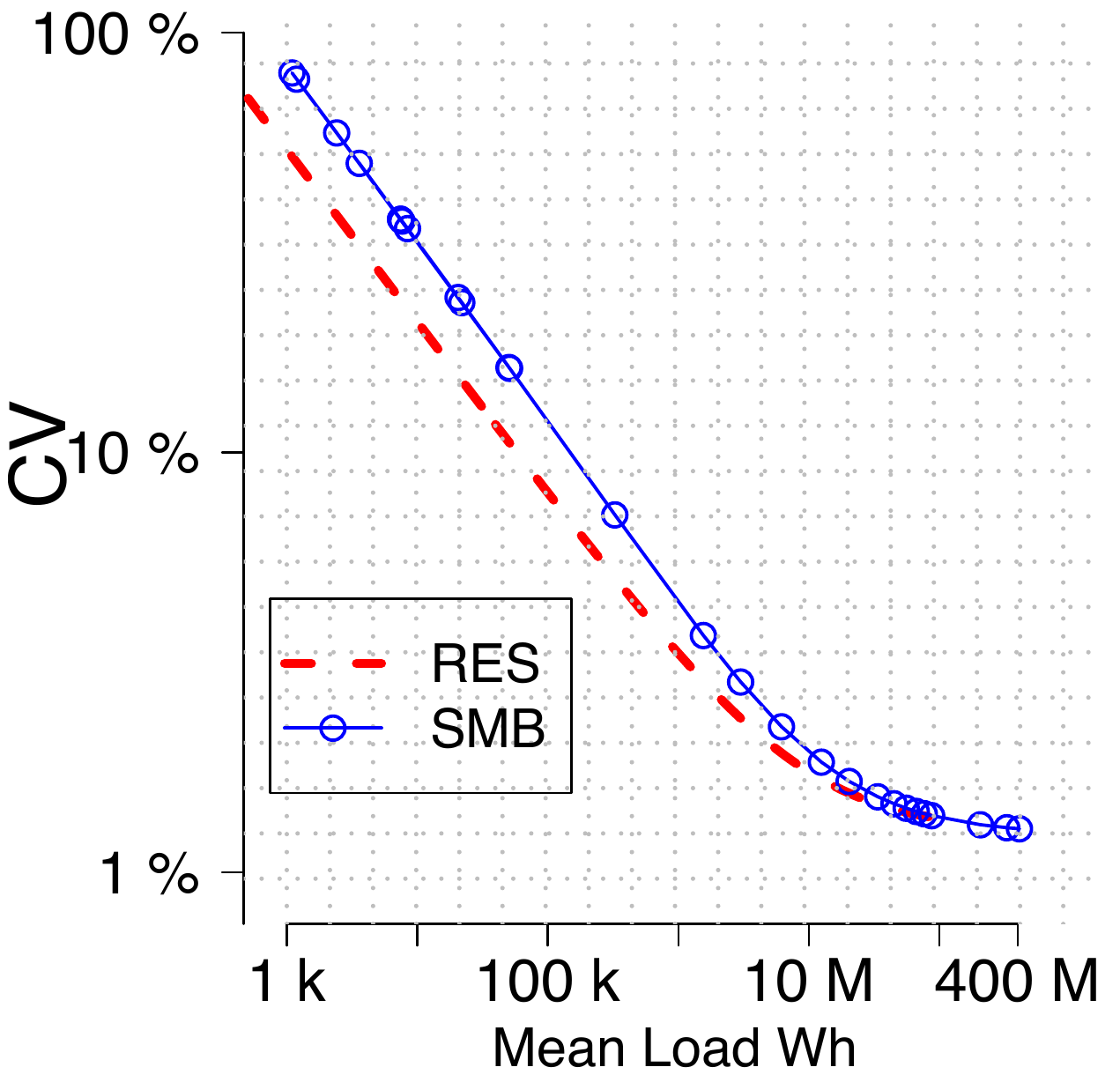} 
}  
\hspace{-5mm}
\subfigure[][] {
\label{fig:unv-coastal-m3}  
\includegraphics[scale=0.27]{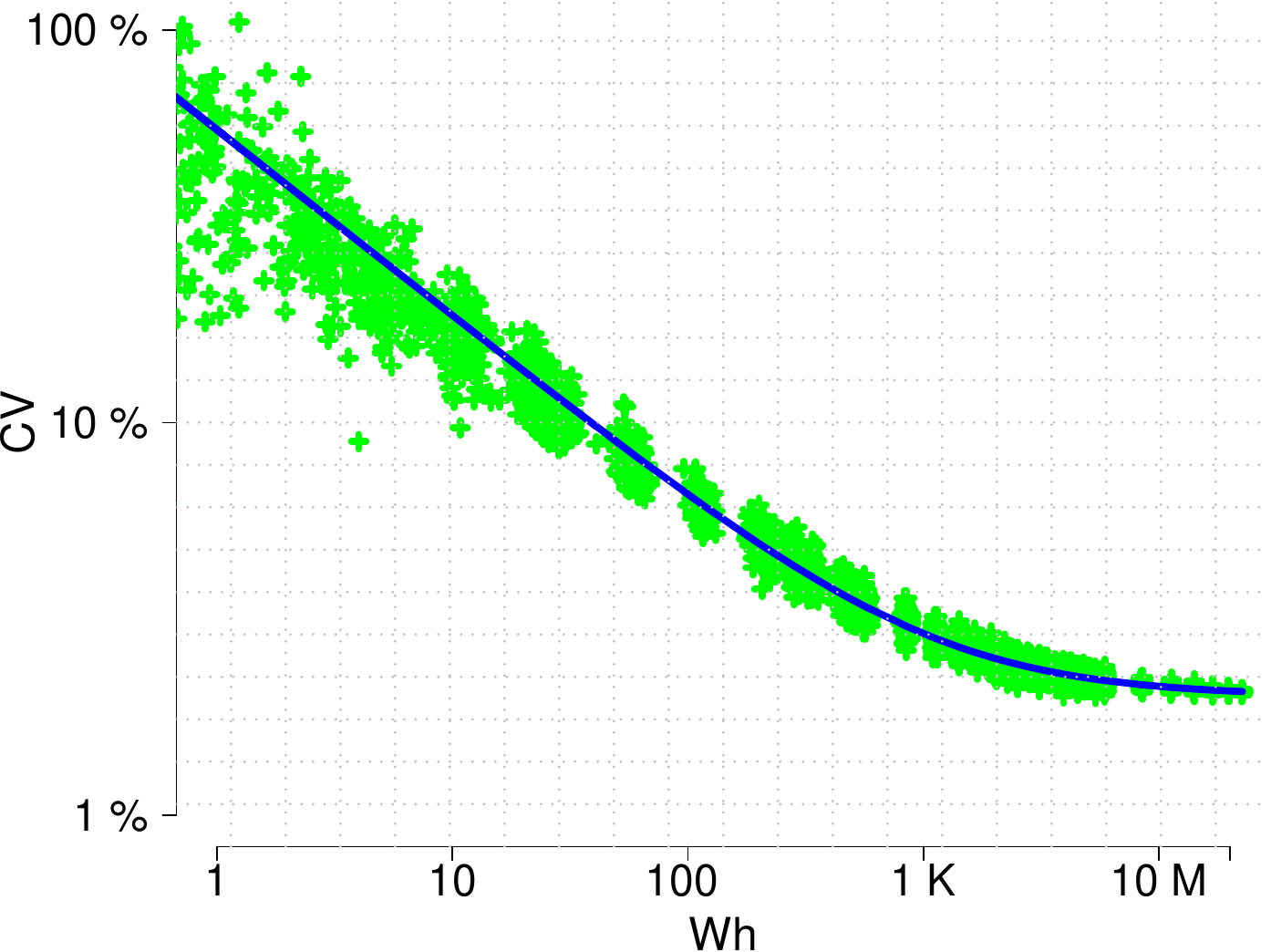} 
}
\hspace{-5mm}
\subfigure[][] { 
\label{fig:unv-inland-m3}
\includegraphics[scale=0.27]{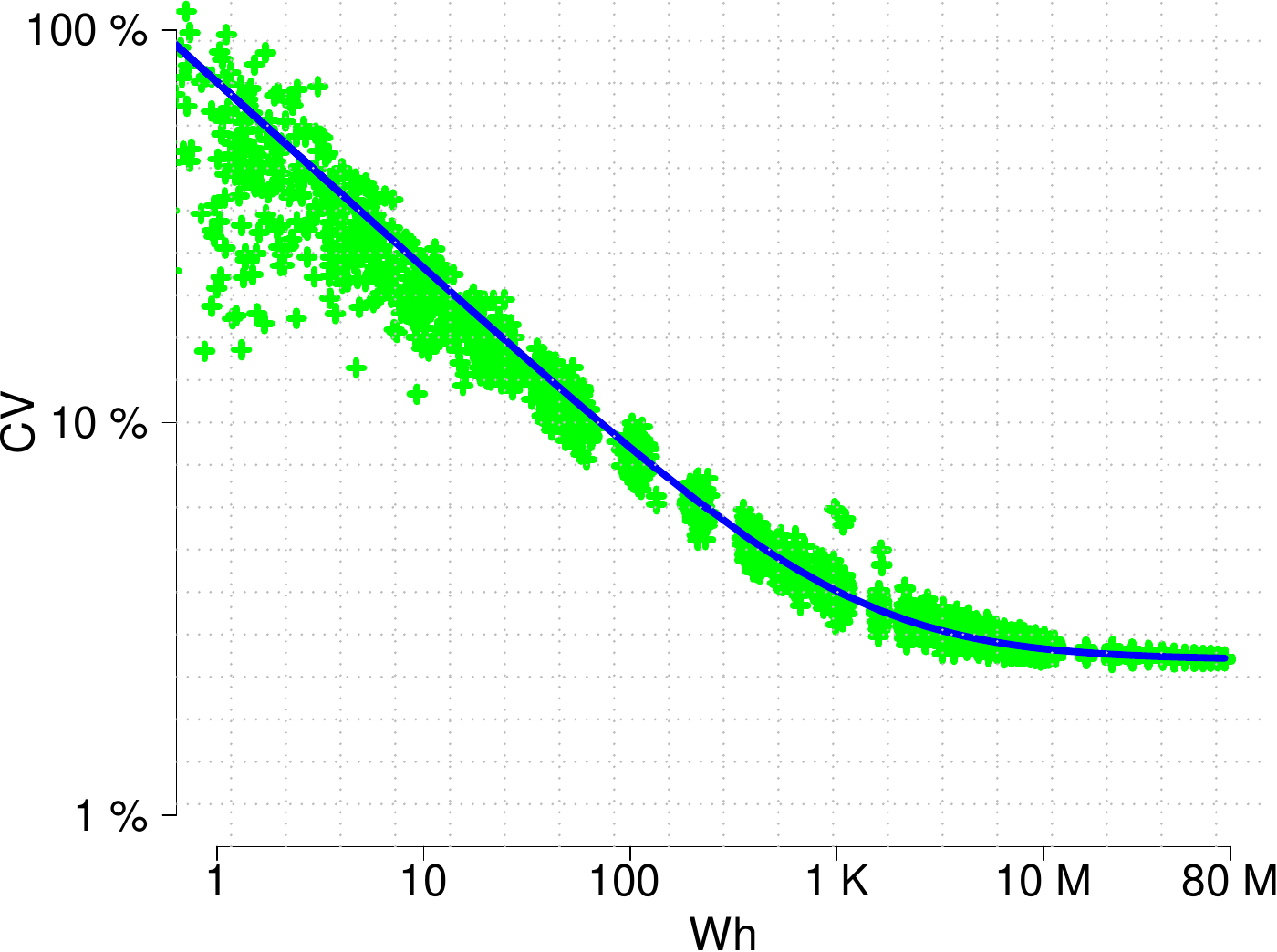}   
}
\caption[]{
Application of aggregation benchmarking procedure for forecaster $\mc{M}_3$
\subref{fig:unv-coastal-m3} coastal population
\subref{fig:unv-coastal-m3} inland population for PG $\&$ E coverage territory.   
}
\label{fig:unv-coastal-inland-m3}
\end{figure}

The benchmarking design in Section \ref{sub-section:Empirical-CV-with-Aggregation-Level} is performed on the PG \& E climate zone populations.
First, the total population is split between the inland and coastal customers.  
Then the aggregates are randomly generated from each climate subpopulation.
The climate subpopulation is separated by the climate zone that each customers zip code falls into.
Following the California PG $\&$ E climate zone designations \cite{CaMap2006}, we define ``coastal populations" as those in climate zones 1, 2, 3, 4 and ``inland populations" as those in climate zones 11, 12, 13, and 14.
The total mean consumption available for the coastal and inland climate zone are 50 MWh and 80 MWh respectively. 
This corresponds to 43,558 coastal customers and 72,558 inland customers.
We then generate aggregates ranging from single users to 40 thousand users for the coastal customers and 65 thousand for the inland customers.
Figure \ref{fig:unv-coastal-inland-m3} shows the aggregation error curve for $\mc{M}_3$ applied to the inland and coastal users.
The results show a similar aggregation error curve under sub-population breakdown.
   
\begin{table}[h]  
\centering
\caption{Aggregation Error Curve Analysis for Coastal Populations (C.P.) and Inland Populations (I.P.)} 
\label{tab:population-model-comparison}
\begin{tabular}{@{}ccccc@{}}
\toprule  
Pop. & Model & p & $\sqrt{\alpha_0}$ & $\sqrt{\alpha_1}$   \\ 
  \cmidrule{1-5}     
C.P. & $\mc{M}_1 $ & 0.91 {\scriptsize(0.86 1.05)} & 63.76 {\scriptsize(58.1 68.6)} & 2.48  {\scriptsize(2.45 2.50)}   \\ 
     ~& $\mc{M}_2 $ & 0.96 {\scriptsize(0.92 1.02)} & 61.14 {\scriptsize(57.1 66.5)} & 1.43   {\scriptsize(1.38 1.49)}  \\  
     ~& $\mc{M}_3 $ & 0.90 {\scriptsize(0.82 0.93)} & 61.90 {\scriptsize(57.6 66.8)} & 1.39   {\scriptsize(1.34 1.46)}  \\ 
~ & ~ & ~ & ~ & ~  \\  
I.P.  &$\mc{M}_1$ & 0.97 {\scriptsize(0.94 1.01)} & 47.06 {\scriptsize(44.33 50.17)} & 1.94 {\scriptsize(1.90 1.99)}    \\ 
      ~&$\mc{M}_2$ & 0.95 {\scriptsize(0.90 1.03)} & 46.32 {\scriptsize(43.29 50.09)}    & 1.35 {\scriptsize(1.26 1.45)}    \\ 
      ~&$\mc{M}_3$ & 0.94 {\scriptsize(0.89 0.98)} & 46.47 {\scriptsize(42.93 49.95)}    & 1.31 {\scriptsize(1.22 1.40)}    \\                              
\bottomrule
\end{tabular}
\end{table}

Table \ref{tab:population-model-comparison} shows the three linear models applied to each set of aggregates.
The linear models were chosen since they were much less computationally expensive to train and test.
The results indicate that in both climate zones, model $\mc{M}_3$ outperforms the other linear models.
Also, the inland dataset has lower irreducible error in the inland dataset as opposed to the coastal dataset.
This is the case even though the coastal dataset has a higher maximum mean load.  
Since the critical loads are $4.8$ MWh and $6.2$ MWh for inland and coastal populations, there are enough samples in the irreducible regime.

%
%
\subsection{Robustness of the Scaling Law}
\label{Ram's New Section}

\begin{figure}[h]
\centering
\includegraphics[width=0.5\textwidth, height=0.4\textwidth]{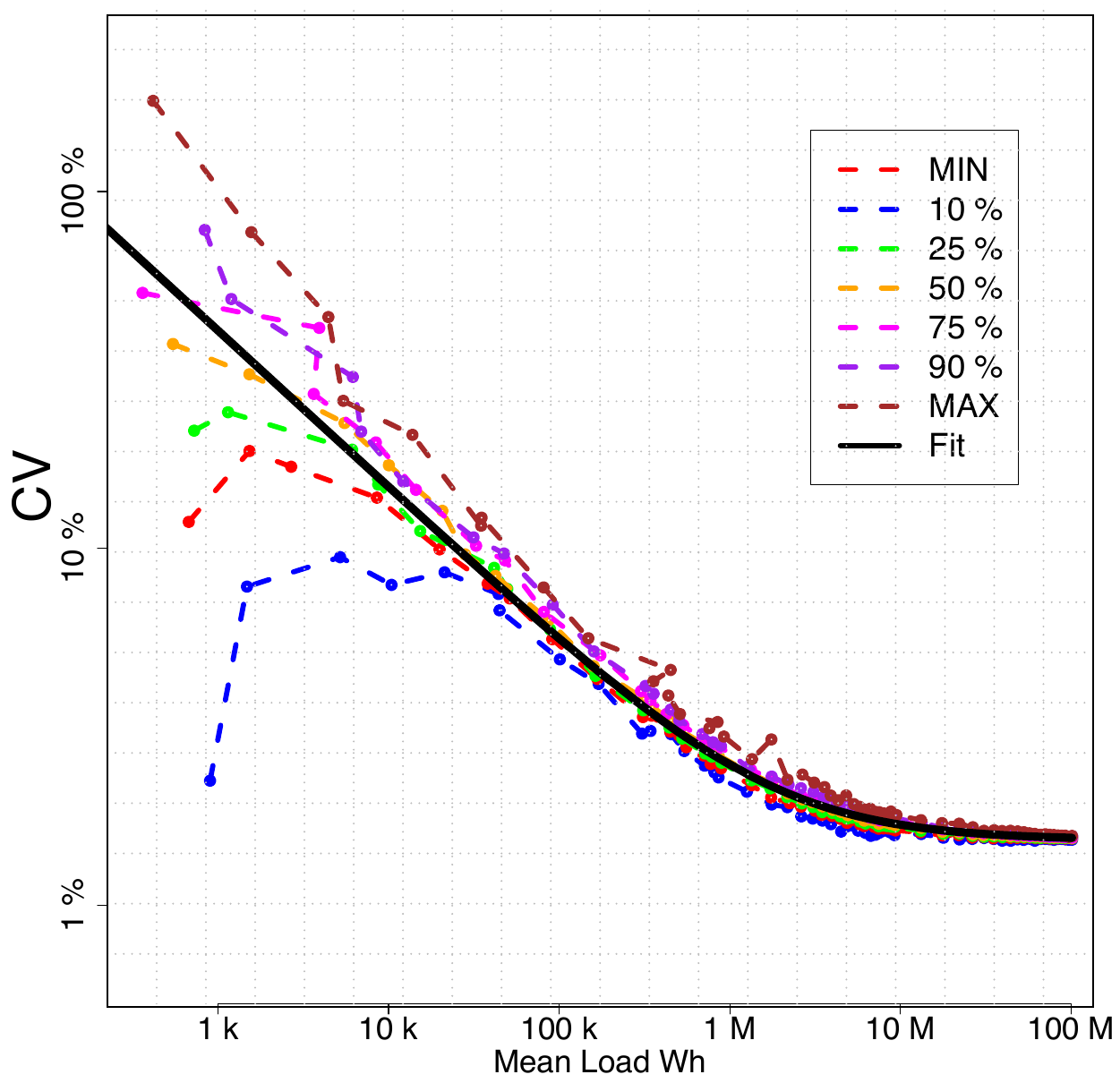}
\vspace{-5mm}
\caption{Comparing quantiles of forecast errors for each aggregation group. The scaling law is robust to the mechanism utilized to generate groups.}
\label{fig:robustness-compare}
\end{figure}

Here we compare the aggregation error curve for the aggregates with best performance and worst performance at each group level. 
The choice is determined by setting a quantile for CV error at each group size for residential data. 
Figure~\ref{fig:robustness-compare} displays the result. 
The scaling law is observed at the different performance quantiles, thus different aggregation mechanisms will obtain results similar to that reported in this paper. 
This indicates that for a general population, the best and the worst sub-groups will have a very close range of relative errors.
For this reason, level of aggregation is an important parameter to having a-priori understanding of the expected coefficient of variation.

%
%
\section{Conclusions and Future Work}
\label{section-conclusion}

This paper introduces the idea of the effect of aggregation on load forecasting.  
We show that forecasting accuracy, as measured in relative error in terms of MAPE and CV improve with larger mean load until a critical load.  
We verify this model with empirical experiments and provide sufficient conditions leading to the observed Aggregation Error Curves introduced in the paper.
It is shown that for various time horizons and models, the proposed model fits the empirical AEC with high accuracy.

Various papers focus on new model formulations to describe individual electricity consumers (e.g. \cite{Ardakanian2011}, \cite{Kolter2012}).  
These models can be utilized to justify a detailed understanding of how aggregate consumption patterns are formed and verified on higher resolution data.  
Moreover, novel ideas can be investigated for aggregate forecasting based on models induced by aggregating this individual consumption models.
The aggregation phenomena is also likely to be observed in other types of forecasting procedures, such as for example day ahead load forecasting, wind forecasting and electric vehicle availability. 
Determining the scaling parameters for these problems is an important task as it can lead to new concepts on the limitations of forecasting big and small aggregates.  

%

\begin{appendix}

\section{Nomenclature Table}
A Table with all terms used in article are provided for ease of description.

\begin{enumerate}
\item[$x(t)$] Consumption Time Series.
\item[$\hat{x}(t)$]                  Forecast Time Series.
\item[$\text{CV}$]                 Coefficient of Variation.
\item[$\text{MAPE}$]            Mean Absolute Percentage Error
\item[$\mc{M}_{k}$]              $k^{th}$ Forecasting Model. 
\item[$W$]                 	    Mean Load of Aggregate. 
\item[$W^{\star}$]                 Critical Load for Aggregation. 
\item[$p$]                              Slope parameter for Empirical Aggregation Error Curve.
\item[$\alpha_0, \alpha_1$]  Model parameters for CV Aggregation Error Curve.
\item[$\beta_0, \beta_1$]      Model parameters for MAPE Aggregation Error Curve.
\item [RES/SMB]                   Residential Customer/Small and Medium Business Customer.
\item[C.P, I.P]                        Coastal/Inland Population.
\end{enumerate}

\section{Aggregation Sizes}
\label{Appendix-Aggregation Sizes}

The following aggregation levels are used in
\begin{align*}
\mb{N}  = \{ &1, 2, 5, 10, 20, 40, 50, 100, 200, 350, 400, 500, 600, 800, 900, 1.0 \times 10^3, 1.5 \times 10^3, 2.0 \times 10^3,  2.5 \times 10^3, \\
	           & 3 \times 10^3, 3.5 \times 10^3, 4 \times 10^3, 4.5 \times 10^3, 5 \times 10^3, 5.5 \times 10^3, 6 \times 10^3,   6.5 \times 10^3, \\
	           & 7 \times 10^3,  7.5 \times 10^3, 8 \times 10^3,  8.5 \times 10^3, 9 \times 10^3,  9.5 \times 10^3, 10  \times 10^3 \\
                   & 10.5 \times 10^3, 15 \times 10^3, 20 \times 10^3,  25 \times 10^3,  30 \times 10^3, 35 \times 10^3,  40 \times 10^3, \\
                   & 45 \times 10^3,  50 \times 10^3, 55 \times 10^3,  60 \times 10^3, 65 \times 10^3, 70 \times 10^3, 75 \times 10^3, \\
                   & 80 \times 10^3, 85 \times 10^3, 90 \times 10^3, 95 \times 10^3, 100 \times 10^3, 105 \times 10^3, 110 \times 10^3, 115 \times 10^3 \}.
\end{align*}

Additionally, the samples per aggregation level are $M = 50$.

\section{Forecasting models}
\label{Appendix-Forecasting-Models}

\subsection{Seasonal Auto Regressive Moving Average (SARMA): $\mc{M}_1 - \mc{M}_3$}
\label{subsection:Seasonal-Auto-Regressive-Models}

SARMA \cite{Box2013Time} predicts the electricity consumption in the next time step as a linear function of prior consumption values and forecast errors. 
Seasonality is considered by including additional predictors at a fixed prior period.
A model SARMA$(p, q){\times}(P, Q)_{s}$ has autoregressive (AR) order $p$ and moving average (MA) order $q$. 
It uses a seasonal component with a cycle of $s$ time steps, with AR order $P$ and MA order $Q$. 
This work considers a restricted class with no MA component so $q = 0$ and $Q =0$. 
The resulting model for the time-series $y(t)$ is
\begin{align}
y[t] = \sum_{k=1}^{p}\theta_{k}y[t-k] + \sum_{k=1}^{P}\phi_{i}y[t-sk] + \epsilon(t). \label{Seasonal-Auto-Regression-Model}
\end{align}
It is usual to assume  $\epsilon(t) \sim N(0, \sigma^2)$ is an independent and identically distributed normal variable.  
The seasonality is set to $s=24$ hours and AR order $P=1$. 
The \textit{adaptive SARMAX} model relearns the parameters $\theta$ and $\phi$ but keeps the parameters $p,P$ and $s$ fixed. 
The SARMA model is applied at each time step by learning the linear model using a pre-set model size.  This constitutes an \textit{adaptive SARMA} model.

\subsection{ Support Vector Regression: $\mc{M}_4$}
\label{subsection:Support-Vector-Regression}
Support Vector Regression (SVR) works by building a non-linear learning method for a training dataset $\left\{(x_1, y_1), \hdots,  (x_N, y_N) \right\}$.  
The training set comprises of $N$ response $y_i$ and predictor $x_i$ pairs.  
The SVR data fitting method solves the following optimization:
\begin{equation*}
\begin{aligned}
& \underset{w, C, \zeta, \zeta^{\star}}{\text{min}}
& & \frac{1}{2}\|w\|^{2} + C\sum_{i=1}^{l} \left( \zeta_i + \zeta^{\star}_i \right) \\
& \text{s.t.} 
& & y_i - w^{T} \Phi(x_i) - b   \le \epsilon + \zeta_{i}  \; i = 1, \ldots, N \\
& & & w^{T} \Phi(x_i) + b - y_1\ge \epsilon + \zeta_{i}^{\star}  \; i = 1, \ldots, N \\ 
& & & \zeta_{i}^{\star} \ge 0,\hspace{5mm} \zeta_{i} \ge 0 \; i = 1, \ldots, N
\end{aligned}
\end{equation*}
Given training predictor $x_i$, there are a given set of kernel functions $\Phi(x_i)$ which map $x_i$ to a high dimensional space.  
The kernel function is fixed and predictions are computed by: $\hat{y}_i=w^{T}\Phi(x_i)+b$.  
The variables $w$, $\Phi$, and $b$ are used to map predictors to response.  
However, fitting the training data only generates vector $w$ and scalar $b$ subject to a set of constraints. 
The SVR will solve for $w$ such that it minimizes the sum of the norm of $w$: $\frac{1}{2}\|w\|^2$ as well a fitting error $C\sum_{i=1}^{l} \left( \zeta_i + \zeta^{\star}_i \right)$.  
The variables $\epsilon$, $\zeta^{\star}$, $\zeta$ quantify the fitting error.  
Any deviation $ |y_i - (w^{T} \Phi(x_i) - b)| \le \epsilon$ incur no penalty.  
However, any deviation outside this dead band ($\zeta^{\star}$, $\zeta$) will incur linear cost yielding $C\sum_{i=1}^{l} \left( \zeta_i + \zeta^{\star}_i \right)$.  
Under this model, the values of $C$, $\epsilon$, kernel function $\Phi$ and additional parameters to $\Phi$ must be specified.  
The support vectors, as well as the constants $C$ and $\epsilon$ are learned adaptively in each training round.
In the training data, we use $3/4$ of the data for training, and $1/4$ for validation of the support vector and constants.
We should note that this method proves computationally expensive but outperforms statically trained models and SARMAX models usually.
For a given kernel function, cross validation is performed to determine the parameters specific to the kernel.  
Then using a moving window we train the SVR and forecast one sample ahead as done in the SARMA model. 
\subsection{ Feed Forward Neural Network (FFNN): $\mc{M}_5$}
\label{subsection:Feed-Forward-Neural-Network}

FFNNs (e.g. \cite{AlShareef2008}) provide a popular alternative to define the nonlinear map between $x(t)$ and  $\mathbf{y}_k(t-1)$ used in the SVR model description.
They are subset of artificial neural networks where neurons with a chosen activation function connect to each other in layers without feedback. 
The number of neurons, layers , choice of activation function and network parameters are learnt from the training set. 
In this model, the training data is used with the $3/4$, $1/4$ split to learn model parameters like in the SVR case.
The following subsection presents work (not included in original publication) for full day ahead forecasting.
Let $x_{d} \in \mathbb{R}^{24}$ be the aggregate daily consumption for days $d = \{1, \hdots, D \}$.
Given previous consumption information $\mathcal{X}_d = \{ x_{1}, \hdots, x_{d} \}$ and daily temperature forecasts $T_d = \{t_1, \hdots, t_{d+1} \}$ the forecaster will output the next day's consumption profile $\hat{x}_{d+1}$.

\subsection{Day Ahead Forecaster: $\mc{M}_6$ }

The forecaster works by predicting the daily total consumption $\hat{p}_d \in \mathbb{R}$ and normalized daily shape pattern $\hat{u}_d \in \mathbb{R}^{24}$ separately.
The final prediction $\hat{x}_{d+1} = \hat{p}_{d+1} \hat{u}_{d+1}$ is the product of each individual forecast.

\textit{Total Power Forecaster}: 
An autoregressive moving average with exogenous input (armax) model is used to forecast the total consumption which is of the form 
\begin{align}
\hat{p}_{d+1} = \sum_{k=d+1-K}^{d} a_{k} p_{k} + \sum_{r=d+1-K}^{d+1} b_{r} t_{r} 
\end{align}
The exogenous input $t_r \in \mathbb{R}$ is the daily mean temperature.
The parameters $a_k,b_r \in \mathbb{R}$ are determined by least squares regression.
A cross validation stage is used to estimate the proper model size $K$.

\textit{Shape Forecaster}: 
A vector ARMAX method is used to forecast the daily shape profile.  
The model is of the form
\begin{align}   
\hat{u}_{d+1} = \sum_{k=d+1-K}^{d} C_{k} u_{k} + \sum_{r=d+1-K}^{d+1} h_{r} t_{r} 
\end{align}
The exogenous input $t_r \in \mathbb{R}^{24}$ contains the mean temperature for each hour.
The parameters $C_{k},h_{r} \in \mathbb{R}^{24 \times 24}$ are real matrices.
These parameters are determined by linear regression given the training data using a least squares formulation.
The model size $K$ is determined in a cross validation stage.

\section{Analytic Model of Aggregation}
\label{appendix-analytical-model-of-aggregation} 

\subsection{Individual Consumption Profile and Forecaster}
\label{appendix-analytical-model-of-aggregation} 

We use the following notation a complete time series vector $\mb{x}$ of length $T$ and a daily profile vector $\mb{x}(d)$ of length $24$, and individual element $x(t)$.
Electricity consumption for individual $n$ is the daily profile vector $\mb{x}(d)$ and is decomposed as $\mb{x}_{n}(d) = \mb{p}_{n}(d) + \epsilon_{n}(d)$ with the following components.
\begin{itemize}
\item $\mb{p}_{n}(d)$ is the daily profile shape for an individual is drawn from a distribution of all load shapes.
This is based off \cite{Kwac2014}, which showed that individual AMI consumption data can be clustered into dictionary of daily shapes.
\item $\epsilon_{n}(d)$ is an additive error.  We make the following assumptions on the first and second order statistics: (1) zero mean $\mbb{E}[ \epsilon_{n}(t) ] = 0$; (2) zero correlated in time $\mbb{E}[ \epsilon_{n}(t) \epsilon_{n}(t+1) ] = 0$; (3) finite population correlation $\mbb{E}[ \epsilon_{n}(t) \epsilon_{n^{\prime}}(t) ] = \gamma$; (4) constant variance $\mbb{E}[ \epsilon_{n}(t) \epsilon_{n}(t) ] = \sigma^2$.  
\end{itemize}

The individual chooses a  daily profile for each day $\mb{p}_{n}(d) \in \mathbb{R}^{T}$ and deviates from it according to $\epsilon_{n}$. 
Therefore a dataset spanning many days is composed of two different stochastic processes: an individual dependent unique shape generation process and a random deviation stochastic process $e_{n}$.  
We do not make assumptions on the shape generating process, since this little work has been done on empirically investigating such a stochastic process.

We use the shorthand for forecast $\mb{\hat{x}}(t+1) = f(\mb{x}, \mc{M})$, which takes as an input the underlying time series to forecast the future horizon, and indexed by the model.
Additionally, we use shorthand $f_{N}(\mc{M}) = f(\sum^N_n \mb{x}, \mc{M})$ to indicate the aggregate forecast under model $\mc{M}$.
In reality, the function will take in all elements up to time $t$.

Now we can compute $CV(W)$, by first computing $CV(N)$ for a finite number of aggregate sizes.
\begin{align}
C(V)  &= \lim_{T \to \infty} CV(N, T) \\
&= \lim_{T \to \infty} \textbf{E}_{\mb{x}} \left[   \left( \frac{ \frac{1}{T} \sum_t^T \left( \sum_n^N x_{n}(t+1) - f_N(\mc{M}) \right)^2  }{   \left( \frac{1}{T}  \sum_t^T  \sum_n^N  x_{n}(t) \right)^2  } \right)^{-\frac{1}{2}} \right]  \\                   
&= \lim_{T \to \infty} \textbf{E}_{\mb{x}} \left[   \left( \frac{ \frac{1}{T} \sum_t^T \left( \sum_n^N x_{n}(t+1) - f_N(\mc{M}) \right)^2  }{   \left( \frac{1}{T}  \sum_t^T  \sum_n^N  x_{n}(t) \right)^2  } \right)^{-\frac{1}{2}} \right]  \\                   
&= \lim_{T \to \infty} \textbf{E}_{\mb{x}} \left[   \left( \frac{ \frac{1}{T} \sum_t^T \left( \sum_n^N p_{n}(t+1) + \epsilon_{n}(t+1)- f_N(\mc{M}) \right)^2  }{  \left( \frac{1}{T} \sum_t^T  \sum_n^N  x_{n}(t) \right)^2 } \right)^{-\frac{1}{2}} \right]  \\                   
&= \lim_{T \to \infty} \textbf{E}_{\mb{x}}  \Biggl[  \Biggl( \frac{ \frac{ 1 }{T} \sum_t^T \left( \sum_n^N p_{n}(t+1) - f_N(\mc{M}) \right)^2 +  \frac{ 2 }{T} \sum_t^T  \left( \sum_n^N \epsilon_{n}(t+1) \right)\left( \sum_n^N p_{n}(t+1) - f_N(\mc{M}) \right)    + \frac{1}{T} \sum_t^T   \left( \sum_n^N \epsilon_{n}(t+1) \right)^2 }{   \left( \frac{1}{T} \sum_t^T  \sum_n^N  x_{n}(t) \right)^2 } \Biggr)^{-\frac{1}{2}}  \Biggr]                    
\end{align}
Here we merely expand the mean squared error term using the decomposition of the time series into the shape profile and additive error term.
We can now analyze each term of the numerator and denominator separately.  
The denominator term is simply:  
\begin{align}
\lim_{T \to \infty}  \frac{1}{T} \sum_t^T  \sum_n^N  x_{n}(t) = \lim_{T \to \infty}  \mc{N}\left( \mu  N, \frac{\omega  N}{T} \right) \xrightarrow[]{a.s.}   \mu  N
\end{align}
This uses the fact that each mean consumption vector is drawn from the distribution given in Figure \ref{fig:data-description-b} (with variance $\omega$) so can be approximated by a normal distribution and using the Law of Large Numbers (LLN).

The last term in the numerator is the following:   
\begin{align}
\lim_{T \to \infty}  \frac{1}{T} \sum_t^T  \left( \sum_n^N \epsilon_{n}(t+1) \right)^2 &= \lim_{T \to \infty}  \frac{1}{T}  \sum_t^T  \left( \mc{N}(0, N \sigma^2 + N ^2 \gamma )  \right)^2 		    \label{eq:additive-err-L1} \\
															 &= \lim_{T \to \infty}  \frac{1}{T}  \sum_t^T  (N \sigma^2 + N ^2 \gamma ) \chi^2 					    \label{eq:additive-err-L2} \\
															 &= \lim_{T \to \infty} \mc{N} \left( N \sigma^2 + N ^2 \gamma  , \frac{ N \sigma^2 + N ^2 \gamma }{T}  \right) \label{eq:additive-err-L3} \\
															 &\xrightarrow[]{a.s.} N \sigma^2 + N ^2 \gamma 										              \label{eq:additive-err-L4} 	
\end{align}

We should note that the sum of the error terms result in a variance which grows linearly and quadratically.
If we assume no correlation across the population, this term will only grow linearly, however as we will show the irreducible error from the AEC must come from some non-zero quadratic term.

The second term of the numerator can be eliminated, since the two terms are uncorrelated and therefore, taking $T \rightarrow \infty$ leads to 
\begin{align}
\lim_{T \to \infty}  \frac{2}{T} \sum_t^T   \left( \sum_n^N \epsilon_{n}(t+1) \right)\left( \sum_n^N p_{n}(t+1) - f_N(\mc{M}) \right) = 0 ~~~~~~~~a.s.
\end{align}

The first term of the numerator is the following: 
\begin{align}
\lim_{T \to \infty} \frac{1}{T} \sum_t^T \left( \sum_n^N p_{n}(t+1) - f_{N}(\mc{M}) \right)^2 = N^2 \underbrace{ \lim_{T \to \infty} \frac{1}{T} \sum_t^T \left(  \frac{1}{N}\sum_n^N p_{n}(t+1) - \frac{1}{N} f_{N}(\mc{M}) \right)^2 }_{\text{Population Bias } \delta( \mb{p},~\mc{M} )^2} 
\end{align}

This represents how well a model can match the average population profile of a particular size given a large enough time period.
We refer to it as a bias term because in a large aggregate, it represents how well the normalized profile fits the average population profile.

Analyzing this via the shape generating process leads to:
\begin{align}
 \delta( \mb{p},~\mc{M} )^2 &= \lim_{T \to \infty} \frac{1}{T} \sum_t^T \left(  \frac{1}{N}\sum_n^N p_{n}(t+1) - \frac{1}{N} f_N(\mc{M}) \right)^2     		\label{eq:pop-bias-L1}	\\
					  & =  \lim_{D \to \infty} \frac{1}{D} \sum_d^D \left(  \frac{1}{N}\sum_n^N \mb{p}_{n}(d) - \frac{1}{N} f_N(\mc{M})  \right)^2  		\label{eq:pop-bias-L2}	\\
					  & =  \lim_{D \to \infty} \frac{1}{D} \sum_d^D \left(  \frac{1}{N}\sum_n^N \mb{p}_{n}(d) -  \textbf{E}_{\mb{p}} \left[  \frac{1}{N}\sum_n^N \mb{p}_{n}(d) \right] \right)^2 	+ \lim_{D \to \infty} \frac{1}{D} \sum_d^D \left(  \textbf{E}_{\mb{p}} \left[  \frac{1}{N}\sum_n^N \mb{p}_{n}(d) \right] - \frac{1}{N} f_N(\mc{M}) \right)^2  	\label{eq:pop-bias-L3}	
\end{align}

The first term in  \eqref{eq:pop-bias-L3} represents the variation of the shape profiles around their average values.
This is a vector of length 24 representing the full day long shape.
We can simplify this to:  
 
\begin{align} 
 \lim_{D \to \infty} \frac{1}{D} \sum_d^D \left(  \frac{1}{N}\sum_n^N \mb{p}_{n}(d) -  \textbf{E}_{\mb{p}} \left[  \frac{1}{N}\sum_n^N \mb{p}_{n}(d) \right] \right)^2  &\xrightarrow[]{a.s.} \frac{1}{N^2} \text{trace} \left( \text{cov} \left(  \sum_n^N \mb{p}_{n} \right) \right) \\
 &= \frac{ \kappa }{ N } 
\end{align} 

This term should generally scale linearly with aggregation size, since it it representing how shapes vary around their mean shape.
In \cite{Kwac2014}, the authors report on similar metrics showing how much variation a single customer has in choice of load shapes.
If we assume this is independent across individuals, then the population variation will decrease in relative terms, leading to $\kappa/N$.

The second term in  \eqref{eq:pop-bias-L3} represents the squared bias between the mean forecast and the mean population shape.
If the model has no population bias, $\textbf{E}_{\mb{x}} \left[ \frac{1}{N} f_N(\mc{M}) \right] =  \frac{1}{N}\sum_n^N \mb{p}_{n}(d)$, and the overall term will have $O\left(\frac{1}{N} \right)$ behavior just as before.
This might happen (although not shown here) if each consumer chooses load shapes according to a steady state distribution, then it is imaginable that an unbiased estimator can be constructed given previous information since it will be the weighted sum of load shapes.
However, if such a model does not exist which captures the shape generating process of a customer, 
\begin{align}
\textbf{E}_{\mb{x}} \left[ \frac{1}{N} f_{N}(\mc{M})  \right]  \neq \frac{1}{N}\sum_n^N \mb{p}_{n}(d),
\end{align}
then  \eqref{eq:pop-bias-L3} will always be some positive non-negative value.
\textit{An interpretation of this term is that it captures how a model can fit the profile generating process of the population.}
This also captures why some forecasters perform better than others. 
Better forecasters capture the shape generating process better than simpler models which will have some large population bias.
If we assume the bias term is forecasting model specific, then $\delta( \mb{p},~\mc{M} ) = \delta(\mc{M})$.

Combining these terms we have the following:
\begin{align*}
CV(N)  &= \lim_{T \to \infty}  CV(N, T) \\
	   &= \textbf{E}_{\mb{p}} \left[ \sqrt{ \frac{N^2 \delta(\mc{M} ) + N (\kappa + \sigma^2) + N ^2 \gamma }{ N^2 \mu^2 } } \right] \\
	   &= \textbf{E}_{\mb{p}} \left[ \sqrt{ \frac{ \delta(\mc{M} ) + \gamma }{ \mu } +  \frac{\kappa + \sigma^2}{N}  } \right].  \\
\end{align*}
If we assume $\delta( \mb{p},~\mc{M} ) = \delta(\mc{M})$ given a large $\mb{p}$, where the population bias depends only on the nature of the forecasting method to learn the shape process, we have the result:
\begin{align}
CV(N) &= \sqrt{ \frac{ \delta(\mc{M}) + \gamma }{ \mu } +  \frac{\kappa + \sigma^2}{\mu N}  } 
\end{align}  
Finally, from our large sample limit, $W = \mu N$, leading to the desired result.
%
%
\subsection{Discussion}

Notice that in the previous section, much of what we did was use an an intuitive load shape model for loads, and impose mild conditions on the generating process and additive error such that they lead to the linear and quadratic growth.
Given only this model formulation, a priori, it is difficult to arrive at any satisfactory conclusion of wether any of the terms should actually be non-zero.
However, given that the empirical AECs show a non-zero irreducible error, we are forced to accept one of the three possibilities:
(1) there is non-zero correlation between additive errors; (2) there is a finite bias term between a forecasting procedure the mean load shape; (3) both (1) and (2) are true.

Under this model, the term $\alpha_0 = \frac{\kappa + \sigma^2}{\mu N}$ and is therefore independent of the forecasting procedure. 
This is somewhat controversial since more forecasting work at small aggregates are basically swamped by this noise.

\section{Scaling Law Parameter Estimation}
\label{appendix-parameter-fit-NLLS}

To estimate the parameters to model \eqref{eq-mean-CV-W}, we assume the following observation model:
\begin{align}
CV(W) = \sqrt{ \frac{\alpha_0}{W^p} + \alpha_1 } + \epsilon
\end{align}     
where $\epsilon$ is a zero mean perturbation.
This is recovered from the AEC curve by transforming the observations $CV(W) \rightarrow CV^2(W)$ and performing a linear regression on: 
\begin{align}
CV^2(W) =  \frac{\alpha_0}{W^p} + \alpha_1  + \epsilon^{\prime} \label{eq-NLLS}
\end{align} 
where now $\epsilon^{\prime}$ is a transformed error term.
Given the regression solution and the estimated $\text{MSE}(p)$, the exponent fit is estimated via $p^{\star} \in \underset{p \in [\underline{p}, \overline{p} ] }{\arg\min}~\text{MSE}(p)$.

\begin{figure}[h]
\centering  
\subfigure[][]{
\includegraphics[scale=0.2]{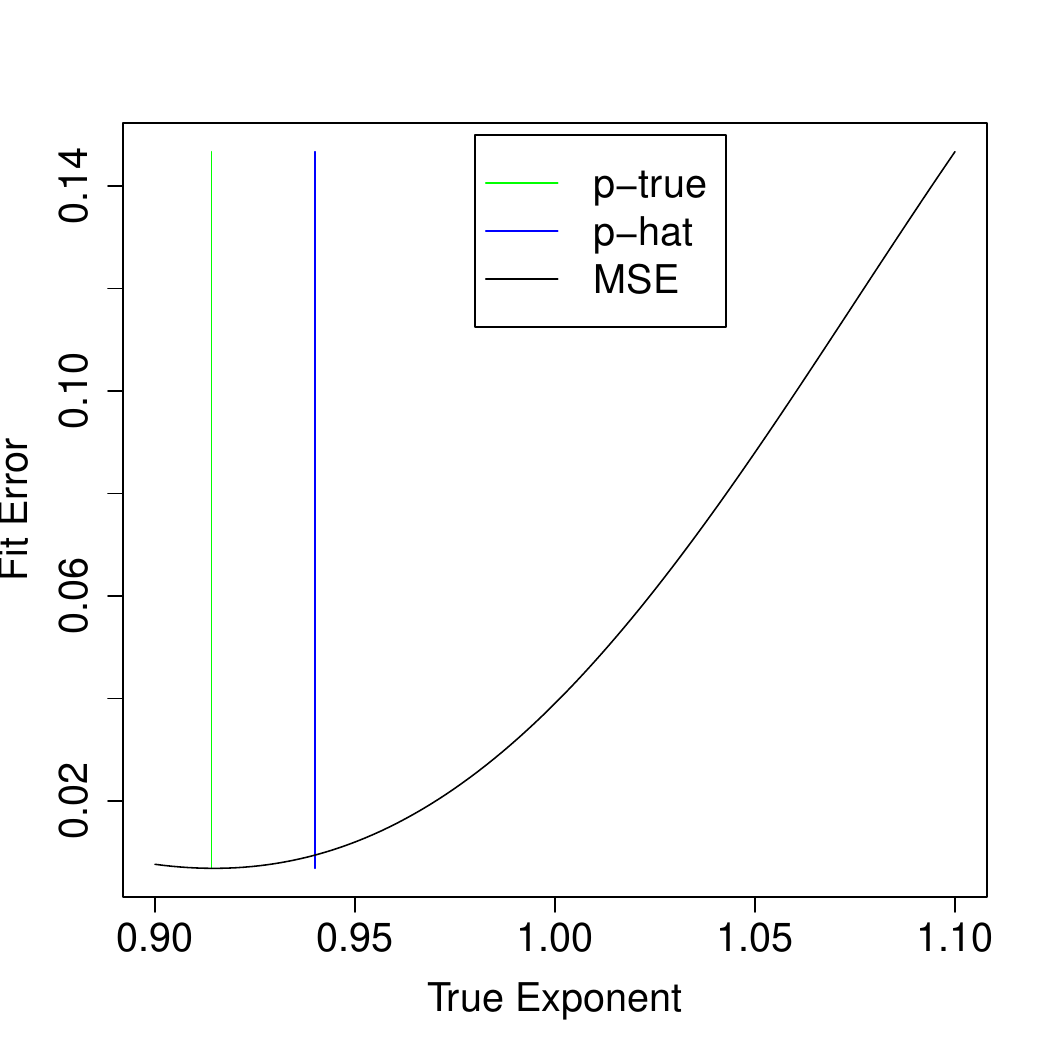}   
\label{fig-parameter-reconstruction-example}  
}  
\subfigure[][]{
\includegraphics[scale=0.2]{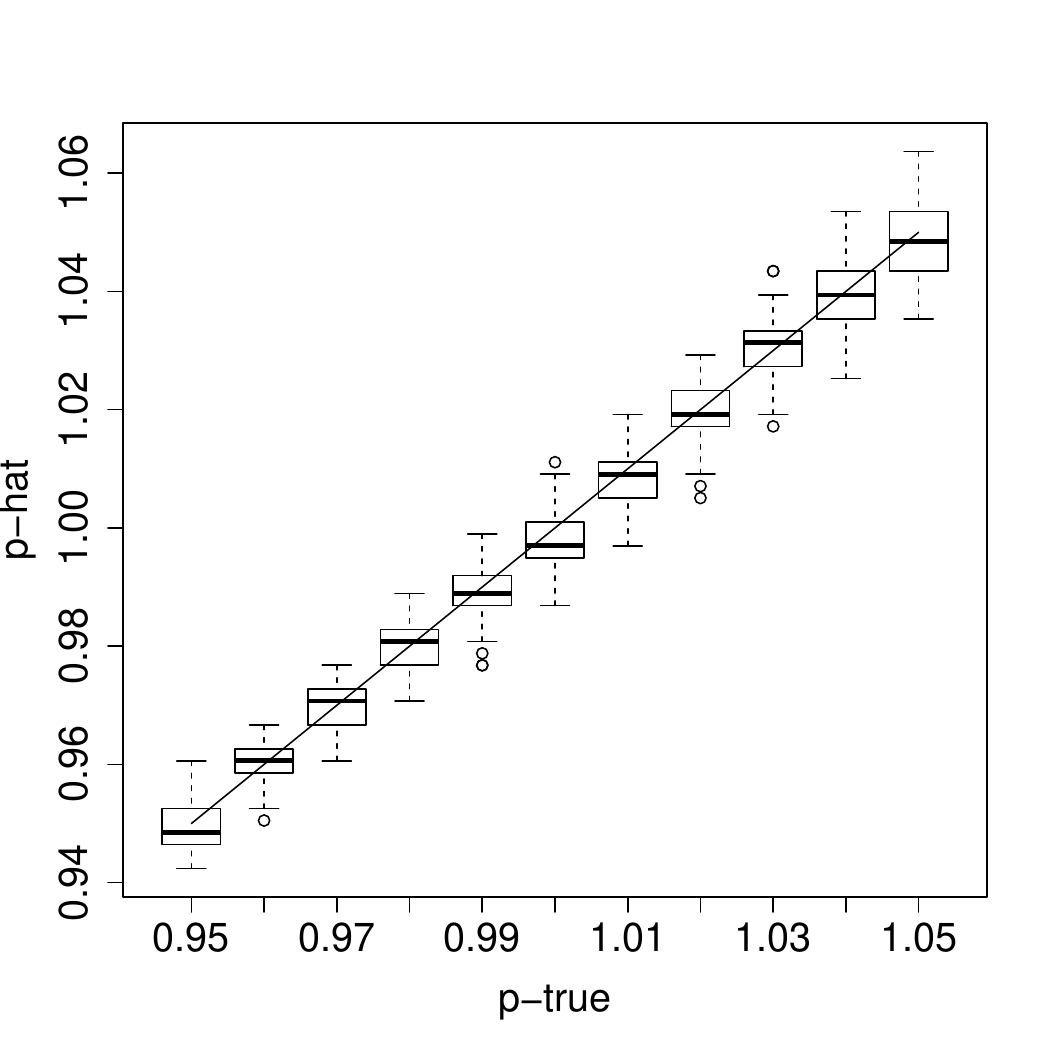}     
\label{fig-p-reconstruction-boxplot}
}  
\subfigure[][]{
\includegraphics[scale=0.2]{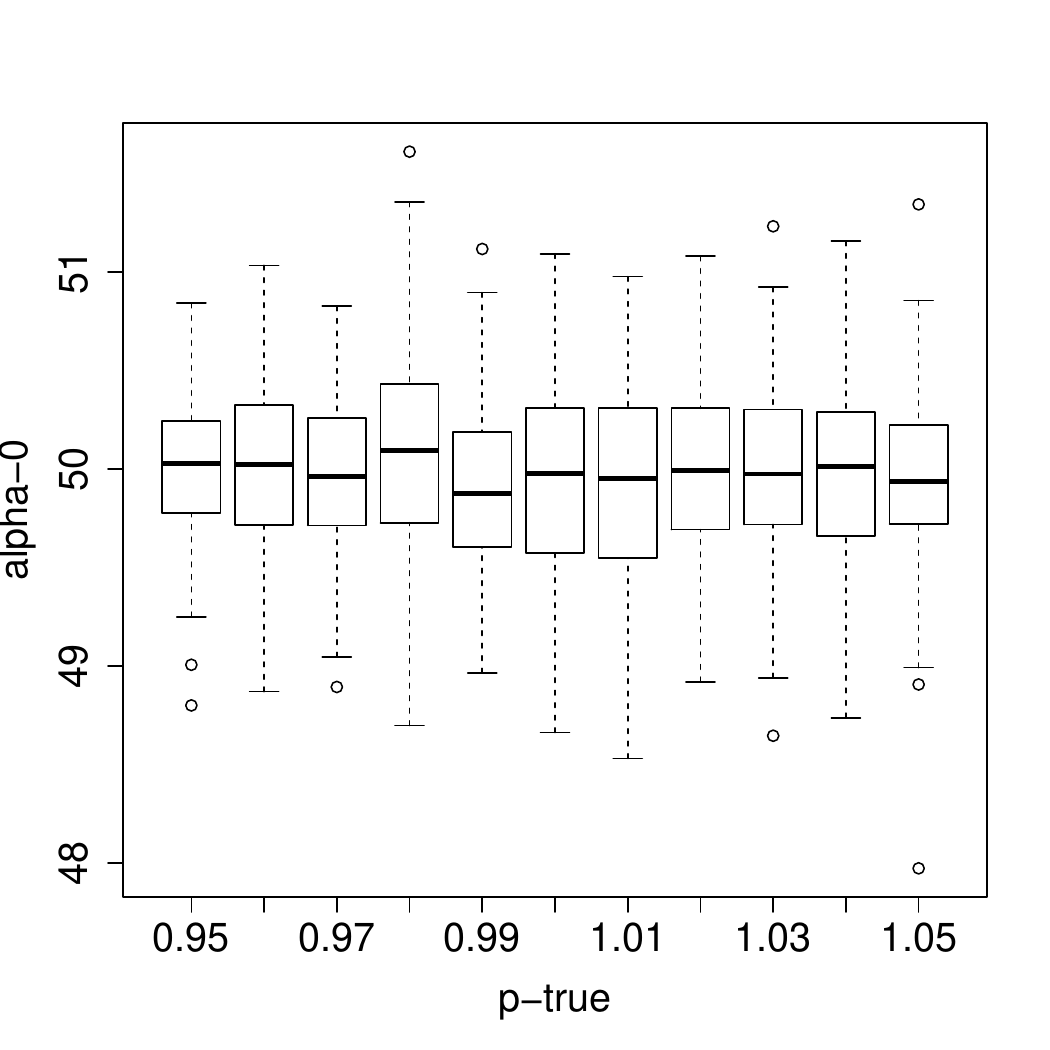}   
\label{fig-a0-reconstruction-boxplot}
}
\subfigure[][]{
\includegraphics[scale=0.2]{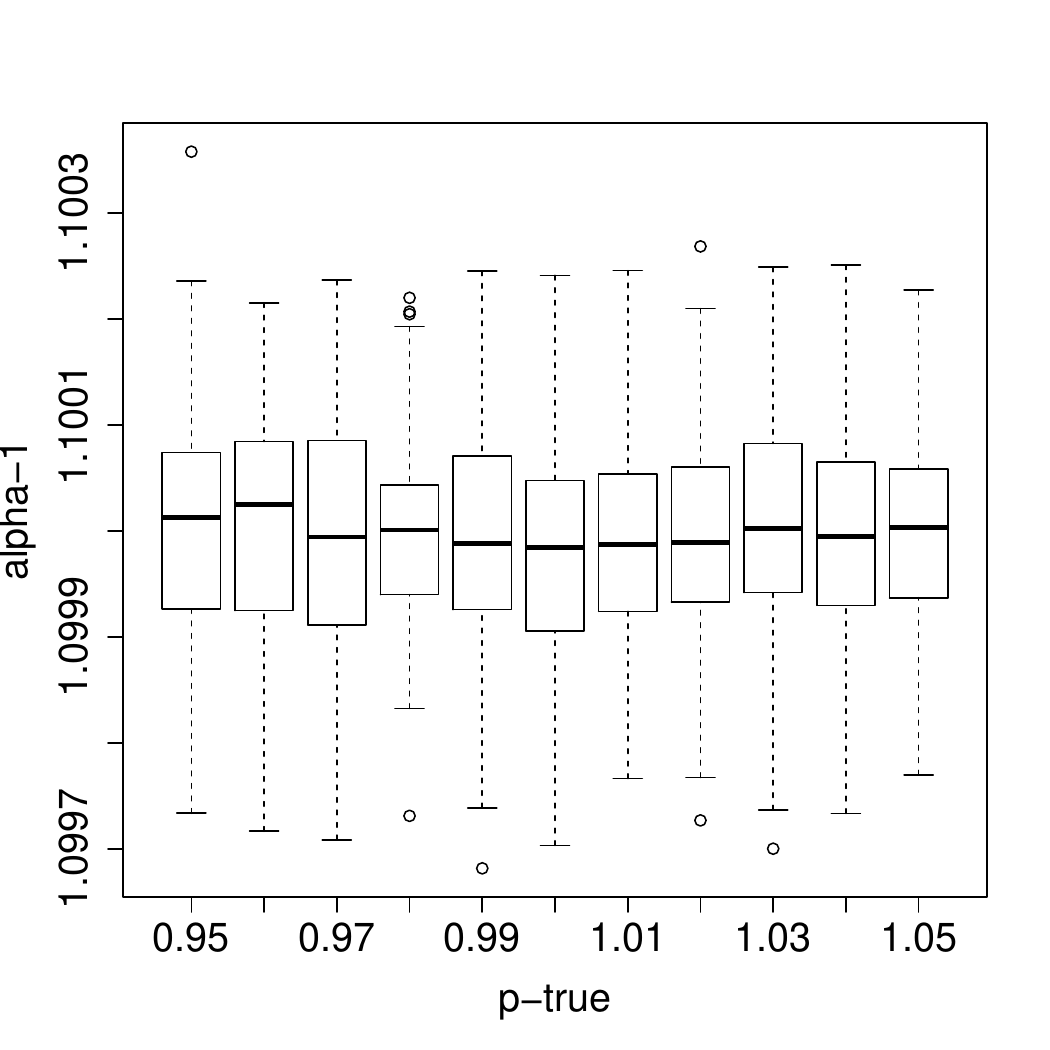}   
\label{fig-a1-reconstruction-boxplot}
}     
\caption[]{  
\subref{fig-parameter-reconstruction-example} Example of fitting the load exponent $p$ in simulated dataset, with $p_{true} = 0.94$ and $\hat{p} = 0.968$.
   Boxplot of estimated \subref{fig-p-reconstruction-boxplot}$\hat{p}$, \subref{fig-a0-reconstruction-boxplot} $\alpha_0$ and \subref{fig-a1-reconstruction-boxplot} $\alpha_1$ values for each Monte Carlo test along with the overall average over range of $p_{true}$.
}
\end{figure}

An important issue is that of possible bias or high variance in estimating the parameter $p$, $\alpha_0$ and $\alpha_1$.
In simulation with known values of $p$ and the estimation procedure \eqref{eq-NLLS}, an example of $\text{MSE}(p)$, $\hat{p}$ and $p_{true}$ are shown in Figure \ref{fig-parameter-reconstruction-example}.
The results show that for a large sample size $50 \times 56 \approx 2500$ there is some error in the estimate.
Regardless over $100$ test runs under a single $p_{true}$, the estimator converges to the correct value for all $p_{true}$ as shown in Figure \ref{fig-p-reconstruction-boxplot}.
Likewise, in estimating the regression coefficients, with $\alpha_{0, true} = 50$ and $\alpha_{1, true} = 1.1$ the non-linear regression procedure recovers the underlying parameters without significant bias as shown in Figure \ref{fig-a0-reconstruction-boxplot}, \ref{fig-a1-reconstruction-boxplot}.
 
\end{appendix}  

\bibliographystyle{elsarticle-num}
\bibliography{tps_bib}

\end{document}